\documentclass[%
 a4paper,
 reprint,
superscriptaddress,
 amsmath,amssymb, amsfonts,
 aps,longbibliography,prx
]{revtex4-2}

\usepackage{graphicx}
\usepackage{dcolumn}
\usepackage{color}
\usepackage{bm}
\usepackage{hyperref}
\usepackage{lipsum}
\usepackage[caption = false]{subfig}

\newcommand{\Rmnum}[1]{\expandafter\@slowromancap\romannumeral  #1@}

\newcommand{\Gb}{ {\bm G } }

\newcommand{\Sigmab}{ {\bm \Sigma } }
\newcommand{\Gammab}{ {\bm \Gamma } }

\newcommand{\Tr}{\text{Tr}}

\usepackage{bbold}

\newcommand{\MB}{\text{MB}}

\begin{document}

\preprint{}

\title{Entanglement and thermokinetic uncertainty relations in coherent mesoscopic transport}

\author{Kacper Prech}
\email{kacper.prech@unibas.ch}
\affiliation{Department of Physics, University of Basel, Klingelbergstrasse 82, 4056 Basel, Switzerland}
\author{Philip Johansson}
\author{Elias Nyholm}
\affiliation{Physics Department, Lund University, Box 118, 22100 Lund, Sweden}
\author{Gabriel T. Landi}
\affiliation{Instituto de F\'{i}sica da Universidade de \~{S}ao Paulo, 05314-970 \~{S}ao Paulo, Brazil}
\author{Claudio Verdozzi}
\affiliation{Physics Department and ETSF, Lund University, PO Box 118, 221 00 Lund, Sweden}
\author{Peter Samuelsson}
\affiliation{Physics Department and NanoLund, Lund University, Box 118, 22100 Lund, Sweden}
\author{Patrick P. Potts}
\affiliation{Department of Physics, University of Basel, Klingelbergstrasse 82, 4056 Basel, Switzerland}


\begin{abstract}

A deeper understanding of the differences between quantum and classical dynamics promises great potential for emerging technologies. Nevertheless, some aspects remain poorly understood, particularly concerning the role of quantum coherence in open quantum systems. On the one hand, coherence leads to entanglement and even nonlocality. On the other, it may lead to a suppression of fluctuations, causing violations of thermokinetic uncertainty relations (TUR and KUR) that are valid for classical processes. 
These represent two different manifestations of coherence, one depending only on the state of the system (static) and one depending on two-time correlation functions (dynamical).
Here we employ these manifestations of coherence to determine when mesoscopic quantum transport can be captured by a classical model based on stochastic jumps, and when such a model breaks down, implying nonclassical behavior. To this end, we focus on a minimal model of a double quantum dot coupled to two thermal reservoirs. In this system, quantum tunneling induces Rabi oscillations and results in both entanglement and nonlocality, as well as TUR and KUR violations. These effects, which describe the breakdown of a classical description, are accompanied by a peak in coherence. Our results provide guiding principles for the design of out-of-equilibrium devices that exhibit nonclassical behavior.

\end{abstract}

\maketitle


\section{Introduction}
\label{sec:Introduction}


The inability of classical physics to account for experimental observations on the atomic scale resulted in the emergence of quantum mechanics in the last century~\cite{Jammer}. Quantum physics exhibits behavior that may be strikingly different from that of classical physics. One fundamentally distinguishing feature is quantum coherence, i.e., the ability to maintain a superposition of different quantum states. Coherence can be understood as a resource in quantum information processing~\cite{Streltsov_2017}, to realize tasks such as quantum computation or quantum metrology. Coherence also plays an important role in the thermodynamics of quantum systems and influences fluctuations of work and entropy production~\cite{Scandi_2020, Latune_2020, Tajima_2021}.

\begin{figure}[t]
    \centering
    \includegraphics[width=0.97\columnwidth]{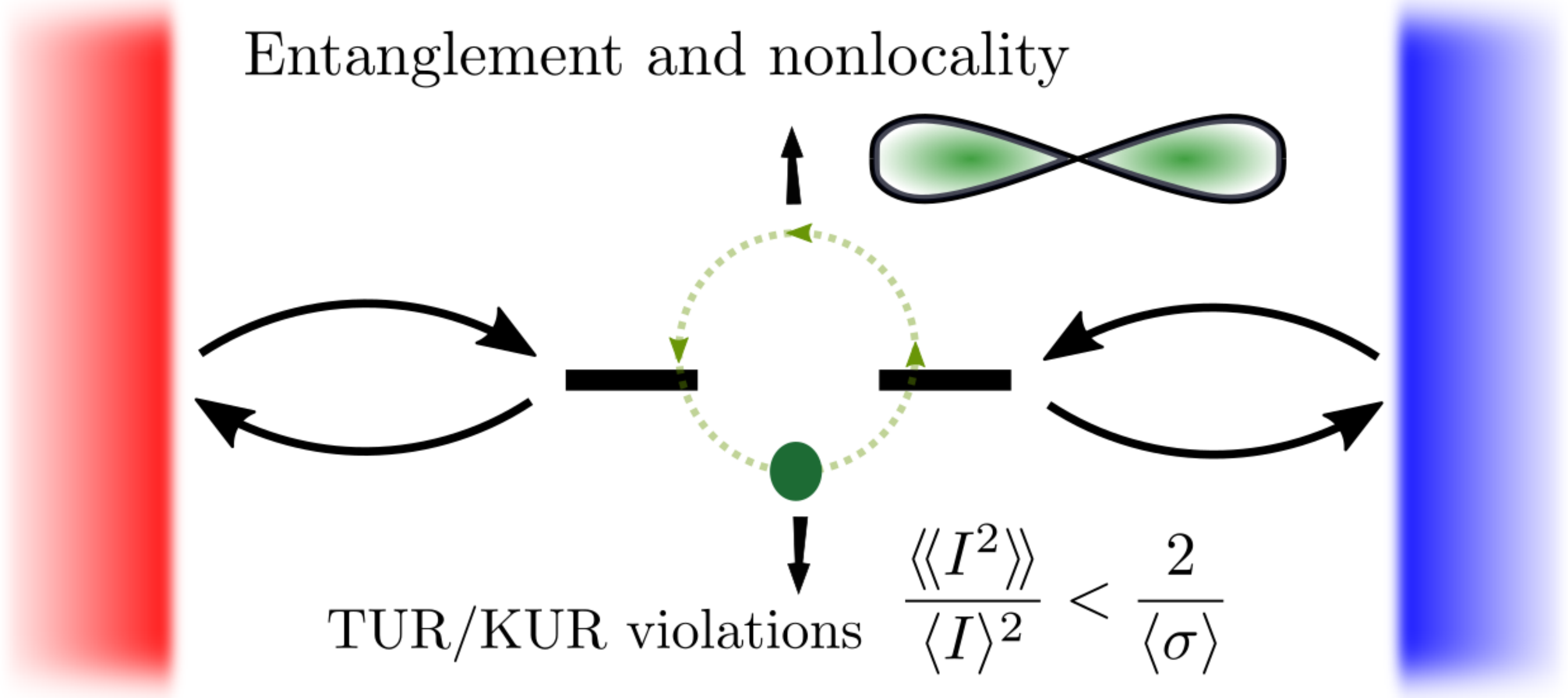}
    \captionsetup{justification=justified, singlelinecheck=false}
    \caption{Manifestations of coherent tunneling in mesoscopic transport. The buildup of quantum coherence, caused by Rabi oscillations, may result in entanglement and even nonlocality. 
    In addition, coherent tunneling may also result in a suppression of fluctuations, leading to violations of thermokinetic uncertainty relations. 
    }
    \label{fig:Summary of results}
\end{figure}

Despite the enormous interest in quantum advantages for technological applications, the difference between quantum and classical behavior has not been fully understood yet. One particular instance is that of quantum transport and thermodynamics in open quantum systems.
Such systems may usually be described by multiple coupled sites, in contact with thermal reservoirs at different temperatures and/or chemical potentials (Fig.~\ref{fig:Summary of results}). After a long time elapses, this system will reach a nonequilibrium steady state (NESS), characterized by currents flowing between the reservoirs. This setup forms the basis for a variety of devices and applications, including thermoelectrics~\cite{benenti_2017,Pekola}, autonomous engines~\cite{Mitchison}, and nanoscale sensors~\cite{Degen}.

Transport through such a system is enabled by quantum tunneling between sites. However, under certain circumstances, a classical model based on stochastic jumps may fully captures the nonequilibrium behavior. To tap into the potential provided by quantum coherence, it is thus of crucial importance to determine when a classical model is adequate to describe an out-of-equilibrium quantum system and when it breaks down. 
To determine the breakdown of a classical description, we consider two manifestations of coherence:
entanglement in the NESS~\cite{Brask_2015,Brask_2022,Khandelwal_2020, Man_2019, Wang_2019} and violations of thermokinetic uncertainty relations~\cite{Kiesslich_2006, Ptaszynski_2018, Agarwalla_2018, Kalaee_2021, Cangemi_2020, Rignon_2021, Menczel_2021, Liu_2019}.


Entanglement~\cite{Horodecki_2009} is an exceptionally strong form of correlation that may result in nonlocal behavior~\cite{Brunner_2014}, which may have an important technological impact as it can be exploited for, e.g., quantum cryptography~\cite{Nielsen,Renner2022} or to provide advantages in sensing applications~\cite{Giovannetti2004,Giovannetti2006,Escher2011}.
Entanglement, though, depends solely on the quantum state of the system at a given time; i.e., it is a static quantity.

The second manifestation is provided by violations of the so-called thermodynamic uncertainty relations (TURs)~\cite{Barato_2015, Gingrich_2016, Horowitz_2019, DiTerlizzi_2018} and kinetic uncertainty relations (KURs)~\cite{DiTerlizzi_2018}, as well as their recently proposed unification, the TKUR~\cite{Vo_2022}. 
These are bounds on the signal-to-noise ratio of currents in classical systems. 
In quantum-coherent systems, these bounds no longer need to be respected, so that TUR or KUR violations can be interpreted as manifestations of quantum coherence.
In contrast to entanglement, the TUR and KUR bounds depend on two-time correlations.
These two types of manifestations of coherence are therefore complementary: while entanglement can be regarded as a static manifestation of coherence, TUR and KUR violations are dynamical in nature.


Here we focus on a minimal model, containing just two sites and two reservoirs (see Fig.~\ref{fig:Summary of results}). In this setup, both NESS entanglement~\cite{Brask_2015,Brask_2022,Khandelwal_2020} as well as TUR violations~\cite{Ptaszynski_2018, Agarwalla_2018} are present. Investigating these manifestations of coherence together, and comparing to a stochastic jump model, allows us determine when quantum coherence results in a behavior that cannot be captured by a classical model. We find that the buildup of coherence due to quantum tunneling can become large enough to generate not only entanglement but even nonlocality, in contrast to previous studies~\cite{Brask_2022}. Furthermore, the Rabi oscillations induced by quantum tunneling result in a reduction of fluctuations that may overcome both the TUR as well as the KUR. To the best of our knowledge, such KUR (and TKUR) violations have not been reported in quantum systems before. Interestingly, both manifestations of coherence occur when the tunnel coupling between the sites is comparable to the system-bath coupling, where the coherence in the NESS shows a local maximum. In addition, entanglement is found for strong tunnel couplings, where coherence exhibits its global maximum. In this regime, however, the dynamics is well captured by a classical jump model that involves the entangled eigenstates of the two-site system. Thermokinetic uncertainty relations may therefore not be violated in this strong-coupling regime.

\begin{figure}[t]
    \centering
    \includegraphics[width=0.97\columnwidth]{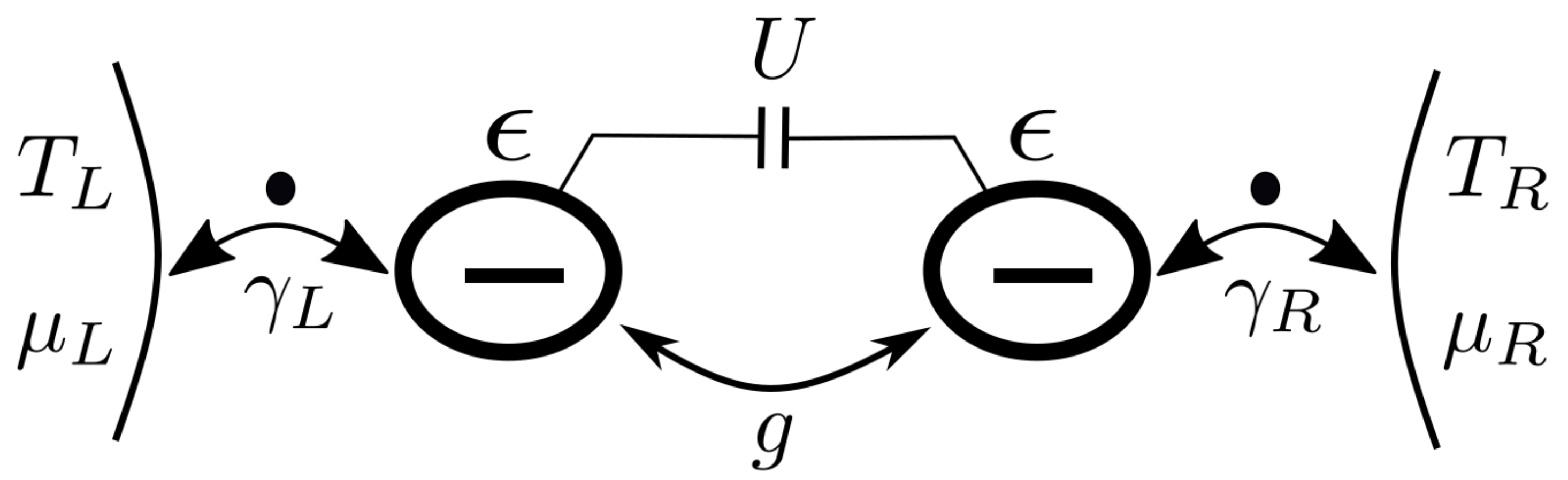}
    \captionsetup{justification=justified, singlelinecheck=false}
    \caption{Sketch of the system. Two spin-polarized quantum dots (QDs) ($\ell = L, R$) with on-site energies $\epsilon$ are tunnel-coupled with strength $g$ and exhibit an interdot Coulomb interaction $U$. Each QD is weakly coupled to a fermionic reservoir with temperature $T_\ell$ and chemical potential $\mu_\ell$. The coupling between system and reservoir is denoted by $\gamma_\ell$.}
    \label{fig:Quantum Dots}
\end{figure}

For concreteness, we consider a double quantum dot, where electrons can hop between two levels. The basic model parameters are outlined in Fig.~\ref{fig:Quantum Dots}. In order to contrast classical models and quantum-coherent models, we employ different quantum master equations~\cite{hofer_2017lvg,Potts_2021} (local vs global) and investigate the effect of interactions.
For a noninteracting system, we benchmark local and global quantum master equations with nonequilibrium Green's functions (NEGFs). We note that we benchmark not only average quantities, depending on the steady state, but also current fluctuations which depend on two-time correlation functions. This further clarifies the regimes of validity of the local and global master equation which have been debated recently \cite{hofer_2017lvg, Gonzalezlvg, Rivas_lvg, Scali_2021}. In the presence of interactions, we employ a thermodynamically consistent semilocal master equation~\cite{Potts_2021, Trushechkin_2021}, as well as NEGFs.  While our results are valid for the two-site model sketched in Figs.~\ref{fig:Summary of results} and \ref{fig:Quantum Dots}, the insights gained in the breakdown of a classical description carry over to more complicated transport scenarios, serving as guiding principles in designing out-of-equilibrium devices that can behave nonclassically. We illustrate this by investigating a chain of three quantum dots.

This paper is organized as follows. 
The system we consider and the models we employ for its description are discussed in Sec.~\ref{sec:System and models}. The manifestations of coherence are introduced in Sec.~\ref{sec:Manifestations}. 
In Sec.~\ref{sec:Non-interacting}, we present the results for noninteracting electrons and in Sec.~\ref{sec:Interacting} we discuss the role of interactions. Section \ref{sec:Extended} contains a discussion on more complicated transport scenarios, including a chain of three quantum dots. Conclusions are provided in Sec.~\ref{sec:Conclusions}.

\section{The system and models}
\label{sec:System and models}

\subsection{Double quantum dot}
We consider two quantum dots (QDs) connected in series, labeled $\ell = L, R$ \cite{van_der_Wiel2002, Sukhorukov_2001, Golovach_2004}. Each QD is weakly coupled to a fermionic reservoir with temperature $T_\ell$ and chemical potential $\mu_\ell$, respectively (Fig.~\ref{fig:Quantum Dots}). 
As shown previously, this system exhibits both manifestations of coherence we are interested in: First, by driving an electron current through the system it is possible to generate an entangled state \cite{Brask_2015}. Second, fluctuations of the same current may be suppressed below the classical TUR bound \cite{Ptaszynski_2018}. We consider spin-polarized electrons, such that at most one electron may reside on each QD. The Hamiltonian of the system reads
\begin{equation}
\label{eq:H}    \hat{H} = \sum_{\ell = L, R} \epsilon \hat{c}_\ell^{\dagger} \hat{c}_\ell  + g \left( \hat{c}_L^{\dagger} \hat{c}_R + \hat{c}_R^{\dagger}\hat{c}_L \right) + U \hat{c}_L^{\dagger} \hat{c}_L \hat{c}_R^{\dagger} \hat{c}_R,
\end{equation}
where $\epsilon$ is the onsite energy of the QDs, $g$ the interdot tunnel coupling, $U$ the Coulomb repulsion between electrons, and $\hat{c}_\ell$ ($\hat{c}_\ell^{\dagger}$) the annihilation (creation) operator of the electron on QD $\ell$, respectively. The effect of unequal onsite energies is considered in Appendix~\ref{App:Non-interacting}.

The effect of the reservoirs is described by multiple methods, including local and global Lindblad master equations as well as NEGFs. 
We now introduce these methods for noninteracting electrons ($U=0$); the interacting case is treated in Sec.~\ref{sec:Interacting}. The strength of the coupling between QD $\ell$ and the corresponding reservoir (assumed to be energy-independent) will be denoted by $\gamma_\ell$ for all methods. Throughout, we consider a voltage bias that is applied symmetrically such that $\mu_L = \epsilon + eV/2$ and $\mu_R = \epsilon - eV/2$.

\subsection{Lindblad master equations}
The main method we use to investigate the dynamics of the double QD is provided by Lindblad master equations. These rely on Born-Markov approximations that for our system are justified when (we set $k_B = \hbar = 1$ throughout)
\begin{equation}
\label{eq:bm}
\gamma_{\ell} \ll \max \{T_{\ell^{'}}, |\epsilon \pm g - \mu_{\ell^{'}}| \},\qquad \text{(Born-Markov)}    
\end{equation}
for any $\ell$ and $ \ell^{'}$ and for both signs. This condition ensures that the relevant bath properties are flat around the Bohr frequencies $\epsilon\pm g$. To ensure thermodynamic consistency, we follow a recently derived framework \cite{Potts_2021, Trushechkin_2021}. Both the local as well as the global master equation that we employ can be written in the form
\begin{equation}
\label{eq:LME}
\frac{d\hat{\rho}}{dt} = -i \left[\hat{H}, \hat{\rho} \right] + \mathcal{L}_L \hat{\rho} + \mathcal{L}_R\hat{\rho} := \mathcal{L} \hat{\rho},
\end{equation}
where
$\mathcal{L}_\ell$ is the Lindblad superoperator that accounts for the dissipation to the bath $\ell$. For the local master equation, the Lindblad superoperators are given by \cite{Potts_2021}
\begin{equation}
\label{eq:Local}
    \mathcal{L}_\ell = n_\ell \gamma_\ell \mathcal{D}[\hat{c}^{\dagger}_\ell] + (1-n_\ell) \gamma_\ell \mathcal{D}[\hat{c}_\ell],
\end{equation}
where
\begin{equation}
\label{eq:Dissipator}
    \mathcal{D}[\hat{O}]\hat{\rho} = \hat{O}\hat{\rho} \hat{O}^{\dagger} - \frac{1}{2} \{\hat{O}^{\dagger}\hat{O}, \hat{\rho} \},
\end{equation}
and $n_\ell := n_\ell(\epsilon)$, with the Fermi-Dirac distribution
\begin{equation}
\label{eq:Fermi}
    n_\ell(\omega) = \frac{1}{\exp{\left(\frac{\omega - \mu_\ell}{ T_\ell}\right)} + 1}.
\end{equation}
When the Born-Markov approximations are justified, the local master equation (ME) may be employed when \cite{Potts_2021}
\begin{equation}
\label{eq: Condition local}
    g \ll \max\{T_\ell, | \epsilon - \mu_\ell|\} \qquad \text{(local ME)}.
\end{equation} 
In analogy to Eq.~\eqref{eq:bm}, this condition ensures that the relevant bath properties are flat across both Bohr frequencies.
In this work, we use the local master equation extensively, as its regime of validity coincides with the regime where both manifestations of coherence appear.

In addition, we make use of the global master equation which relies on the secular approximation \cite{Breuer}. The Lindblad superoperators in the global approach read \cite{Potts_2021}
\begin{equation}
\label{eq:Global}
    \mathcal{L}_\ell = \frac{1}{2} \sum_{s= \pm} \left\{ n_{\ell}^s \gamma_{\ell} \mathcal{D}[\hat{c}_s^{\dagger}] +  (1 - n_{\ell}^s) \gamma_{\ell} \mathcal{D}[\hat{c}_s]\right\},
\end{equation}
where
\begin{equation}
    \epsilon_{\pm} = \epsilon \pm g,\hspace{1.5cm} \hat{c}_\pm = \frac{1}{\sqrt{2}} \left(\hat{c}_R \pm  \hat{c}_L\right),
\end{equation}
and $n_{\ell}^s := n_{\ell}(\epsilon_s)$. The annihilation operators $\hat{c}_\pm$ correspond to the eigenmodes of Hamiltonian \eqref{eq:H}. The secular approximation is justified when~\cite{Breuer}
\begin{equation}
    g \gg \max\{\gamma_L, \gamma_R\} 
    \qquad \text{(global ME).}
\end{equation}

In both Lindblad master equations, the steady-state average current flowing across the system can be computed with the expression
\begin{equation}
\label{eq: Current SS}
    \langle I \rangle = \text{Tr} \left[ \hat{N} \mathcal{L}_L \hat{\rho} \right],
\end{equation}
where 
\begin{equation}
\hat{N} = \hat{c}_L^{\dagger} \hat{c}_L + \hat{c}_R^{\dagger} \hat{c}_R  
\end{equation}
is the operator of the number of electrons in the double quantum dot. 
To determine the current fluctuations, we resort to the method of full counting statistics discussed in Appendix~\ref{App:FCS}.

We note that in both master equations, we neglect a Lamb shift that results in a shift of the on-site energies.

\subsection{Nonequilibrium Green's functions}
For noninteracting electrons, all relevant quantities can be solved exactly using the method of NEGFs. The transmission function $\mathcal{T}(\omega)$, which specifies the transition rate of electrons with energy $\omega$, is a fundamental quantity to investigate transport across the system. For our system it is given by \cite{Sumetskii_1993, Agarwalla_2018}
\begin{equation}
    \label{eq:Transmission function}
    \mathcal{T}(\omega) = \frac{\gamma_L \gamma_R g^2}{|\left(\omega - \epsilon + i\frac{\gamma_L}{2} \right) \left(\omega - \epsilon + i\frac{\gamma_R}{2} \right)-g^2|^2},
\end{equation}
where we assumed the wideband approximation~\cite{Jauho_1994}, i.e., energy independence of the coupling strength $\gamma_\ell$ and omitting any Lamb shift.
The transmission function allows us to compute the average of the current and its fluctuations, given by \cite{Levitov_1993, Levitov_1996, Levitov_2004}
\begin{equation}
    \langle I \rangle = \int_{-\infty}^{\infty} \frac{d\omega}{2 \pi} \mathcal{T}(\omega) \left[n_L(\omega) - n_R(\omega) \right],
\end{equation}
and
\begin{equation}
\begin{split}
    \langle\!\langle I^2 \rangle\!\rangle &= \int_{-\infty}^{\infty} \frac{d\omega}{2 \pi} \mathcal{T}(\omega) \{n_L(\omega) + n_R(\omega) -2 n_L(\omega) n_R(\omega) \\
    &
    - \mathcal{T}(\omega)[n_L(\omega) - n_R(\omega)]^2 \}.
\end{split}
\end{equation}
The elements of the density matrix of the system can be computed from the lesser Green's functions together with Wick's theorem. More information on the NEGF formalism can be found in Appendix~\ref{App:NEGFs}.

\subsection{Classical model}
The last method that we employ is a classical, stochastic model for the double QD system \cite{sprekeler_2004,Kiesslich_2006}. We consider the time evolution of the vector of probabilities $\vec{p} = [p_0, p_L, p_R, p_D]^T$, where entries denote, respectively, the probability that the system is empty, only the left QD is occupied, only the right QD is occupied, and both QDs are occupied. In the classical model, the time evolution of $\vec{p}$ is governed by the rate equation
\begin{equation}
\label{eq:SM}
    \frac{d}{dt} \vec{p} = W \vec{p},
\end{equation}
where the elements $W_{i\neq j}$ of the $4\times4$ matrix $W$ are the transition rates from state $j$ to state $i$, and $W_{ii}=-\sum_{j\neq i} W_{ji}$. The transition rates between states with a different total number of electrons are taken from the local master equation, i.e., $W_{L0} =W_{DR}= n_L \gamma_L$, $W_{0L}=W_{RD}=(1-n_L)\gamma_L$, and similarly for $L\leftrightarrow R$. The transition rate describing hopping between the dots can be obtained from perturbation theory (cf.~Appendix~\ref{App:Stochastic model}) and reads
\begin{equation}
    \label{eq:ratebetweendots}
    W_{LR} = W_{RL} = \frac{4g^2}{\gamma_L + \gamma_R}.
\end{equation}
The average current can be evaluated as
\begin{equation}
    \langle I \rangle = W_{L0} p_0 - W_{0L} p_L + W_{DR} p_R - W_{RD} p_D.
\end{equation}
To compute the fluctuations of the current we resort to full counting statistics (cf. Appendix~\ref{App:FCS}), as in the case of the Lindblad master equations.

\section{Manifestations of coherence}
\label{sec:Manifestations}

\subsection{Coherence}

We study coherence in the quantum state $\hat{\rho}$ of the system with respect to the basis of the particle occupation number of each quantum dot. A local Fock basis $\{|n_L, n_R \rangle := (\hat{c}_L^{\dagger})^{n_L} (\hat{c}_R^{\dagger})^{n_R} |0, 0 \rangle \}$ is constructed such that the first (second) number denotes the occupation of the left (right) QD. Then, the density matrix of the system can be written in the basis $\{|0, 0\rangle,  |1, 0\rangle, |0, 1\rangle, |1, 1\rangle\}$ as a matrix 
\begin{equation}
\label{eq:Density Matrix}
    \hat{\rho} = 
    \begin{pmatrix}
        p_0 & 0 & 0 & 0 \\
        0 &  p_L &  \alpha & 0 \\
        0 &  \alpha^{*} &  p_R & 0 \\
        0 & 0 & 0 &  p_D
    \end{pmatrix}.
\end{equation}
It has a single off-diagonal element $\alpha$, corresponding to a single electron being in a superposition between the left and the right dots. Any other off-diagonal element is strictly zero due to the charge superselection rule \cite{wick_1952,bartlett_2007}. Throughout, we will quantify coherence with the absolute value $|\alpha|$, which is equivalent to one half of the $l_1$-norm of coherence \cite{Baumgratz_2014}. 

\subsection{Entanglement}

The system sketched in Fig.~\ref{fig:Quantum Dots} and its modifications have attracted a significant amount of attention because they may feature entanglement in the steady state \cite{Brask_2015, Brask_2022, Khandelwal_2020, Man_2019, Wang_2019}. This entanglement may arise from two different mechanisms. For tunnel couplings comparable to the system-bath coupling, the entanglement is generated by a current passing through the double dot \cite{Khandelwal_2020}. This scheme to produce nonseparable states can be extended to multiqubit machines \cite{Tavakoli_2020}. It can also be improved by local filtering \cite{Tavakoli_2018}, and it can be mediated by a cavity mode \cite{Tacchino_2018}.

For tunnel couplings much larger than the temperature, only the lowest energy eigenstate is occupied. This state is an entangled state such that the thermal state becomes entangled. In this paper, we are mainly interested in the generation of entanglement by passing a current through the system because this is the regime where we also observe other manifestations of coherence.

We emphasize that the introduced scheme generates mode entanglement between two QDs, where the number of particles plays the role of the degree of freedom \cite{Friis_2016, Dasenbrook_2016, bartlett_2007, Tan_1991}. This is qualitatively different from entanglement between two particles, such as spin entanglement in electronic systems. We note that the entanglement of a single electron~\cite{SingleElectronEnt} has been debated in the literature due to the charge superselection rule. However, following \cite{Dasenbrook_2016}, having two copies of the system allows to overcome this issue and makes the entanglement of a single electron useful.

To quantify the amount of entanglement we use the entanglement measure of concurrence \cite{Hill_1997, Wooters_1998}. For the present system, the concurrence is given by
\begin{equation}
\label{eq:Concurrence}
    \mathcal{C} =  \text{max}\left\{2|\alpha| - 2\sqrt{p_0 p_D}, 0\right\}.
\end{equation}
From this expression, we can already anticipate that Coulomb interactions have a favorable effect on the entanglement, because they reduce $p_D$. This has a twofold effect as it allows for a larger occupation in the single-electron subspace, which may result in a larger $\alpha$.

\subsection{Thermodynamic and kinematic uncertainty relations}

The TUR and  KUR are two different inequalities that bound the signal-to-noise ratio of a current in classical systems. The TUR has been developed in the field of classical stochastic thermodynamics \cite{Seifert_2012, Jarzynski_2011, Bochkov_2013, Mansour_2017, Harris_2007}, which is a framework to describe fluctuations of thermodynamic quantities, such as work or heat, in nanoscale systems. It provides the bound in terms of the average entropy production rate, which quantifies the irreversibility of a process. The TUR can be applied to obtain a trade-off between power production, constancy, and efficiency in stochastic heat engines \cite{Pietzonka_2018} and to infer the efficiency of molecular motors \cite{Pietzonka_2016}.

The KUR was derived using the fluctuation-response inequality for out-of-equilibrium stochastic dynamics \cite{Dechant_2020} and finds applications outside of the thermodynamic settings. In the case of the KUR, the signal-to-noise ratio is bounded by the average dynamical activity, which is a measure of the total rate of transitions between the states of the system. The question whether the KUR can be violated in a quantum system has, to the best of our knowledge, not been explored. We note that in contrast to entropy production, which can be derived using any model, subtleties arise when computing the dynamical activity that features in the KUR with different models. These are discussed below. 

While the TUR provides a tight bound when dissipation is small, the KUR is more relevant for systems that are far from equilibrium. Either of these two different inequalities may thus provide a stronger bound on the fluctuations depending on the situation. As discussed below, these inequalities have also recently been combined into the TKUR \cite{Vo_2022}.

The TUR quantifies a trade-off between an average of a current, $ \langle I\rangle $, its fluctuations $\langle \! \langle I^2 \rangle \! \rangle $, and the total entropy production rate $\langle \sigma \rangle$. It was initially proven in the long-time limit for nonequilibrium systems that follow a time-homogeneous Markovian dynamics and obey local detailed balance. It is given by \cite{Barato_2015, Gingrich_2016}
\begin{equation}
\label{eq:TUR}
    \mathcal{Q}_T := \frac{2 \langle I \rangle^2}{\langle\! \langle I^2 \rangle\! \rangle \langle \sigma \rangle} \leq 1.
\end{equation}
In addition, several modified versions of the TUR have been derived, which include a finite time generalization \cite{Pietzonka_2017, Horowitz_2017}, measurement and feedback scenarios \cite{Potts_2019}, as well as time-dependent periodic \cite{Proesmans_2017} and nonperiodic \cite{Koyuk_2020} driving. In this work, however, we restrict our attention to the TUR given in Eq.~\eqref{eq:TUR}. In the steady-state regime, entropy production comes solely from the dissipation of heat within the reservoirs~\cite{benenti_2017},
\begin{equation}
\label{eq:Entropy production}
    \langle \sigma \rangle = -\frac{ J_L }{T_L} -\frac{ J_R }{T_R},
\end{equation}
where $ J_\ell $ is the steady state heat current from the reservoir $\ell$. How these heat currents can be computed from the different models is discussed in Appendix~\ref{App:Non-interacting}. For reservoirs with equal temperature $T$, the entropy production simplifies to
\begin{equation}
    \langle \sigma \rangle = \frac{(\mu_L - \mu_R) \langle I \rangle }{T} = \frac{eV \langle I \rangle }{T}.
\end{equation}
Moreover, Eq.~\ref{eq:TUR} reduces to 
\begin{equation}
\label{eq:TUR2}
    \mathcal{Q}_T := \frac{2 T \langle I \rangle}{eV\langle\! \langle I^2 \rangle\! \rangle } \leq 1,
\end{equation}
which only depends on the Fano factor $\langle\! \langle I^2 \rangle\! \rangle/\langle I \rangle$.

The KUR provides an alternative bound on the signal-to-noise ratio \cite{DiTerlizzi_2018},
\begin{equation}
    \label{eq:KUR}
    \mathcal{Q}_K := \frac{ \langle I\rangle^2}{\langle\! \langle I^2 \rangle\! \rangle \langle \mathcal{A} \rangle} \leq 1,
\end{equation}
where $\langle \mathcal{A} \rangle$ is the dynamical activity that quantifies the average rate of jumps the system undergoes. For a system described by a classical rate equation, such as Eq.~\eqref{eq:SM}, the dynamical activity is defined as \cite{DiTerlizzi_2018}
\begin{equation}
\label{eq:Dynamical activity stochastic}
    \langle \mathcal{A} \rangle = \sum_{i\neq j} W_{ij}p_j.
\end{equation}
It quantifies the average total rate of jumps between all states of the system. In the quantum model, the process of electrons moving between the left and right QDs is modeled by a coherent evolution, not classical jumps. For this reason, the dynamical activity for Markovian quantum master equations is usually defined such that it only quantifies the rate of jumps induced by Lindblad jump operators \cite{Hasegawa_2020, Vu_2022}. For the local master equation, this would read
\begin{equation}
\label{eq:Dynamical activity quantum}
    \langle \mathcal{A}_q \rangle = \langle \mathcal{A} \rangle -\frac{4g^2}{\gamma_L+\gamma_R}(p_L+p_R),
\end{equation}
resulting in a strictly smaller dynamical activity. Employing $\langle \mathcal{A}_q \rangle$ in the KUR could result in the following problem: In a regime where the classical model and the master equation provide the same probabilities and fluctuations (i.e., the system essentially behaves classically), we could get a violation of the KUR simply because $\langle \mathcal{A}_q \rangle$ does not take into account the transitions between the left and right dots. To avoid such problems, we employ Eq.~\eqref{eq:Dynamical activity stochastic} throughout for the dynamical activity, where the rates $W_{ij}$ are always taken from the classical model \eqref{eq:SM}, and the probabilities $p_j$ are obtained with the respective method that is employed.

\begin{figure*}[t]
        \centering
        \includegraphics[width=1\textwidth]{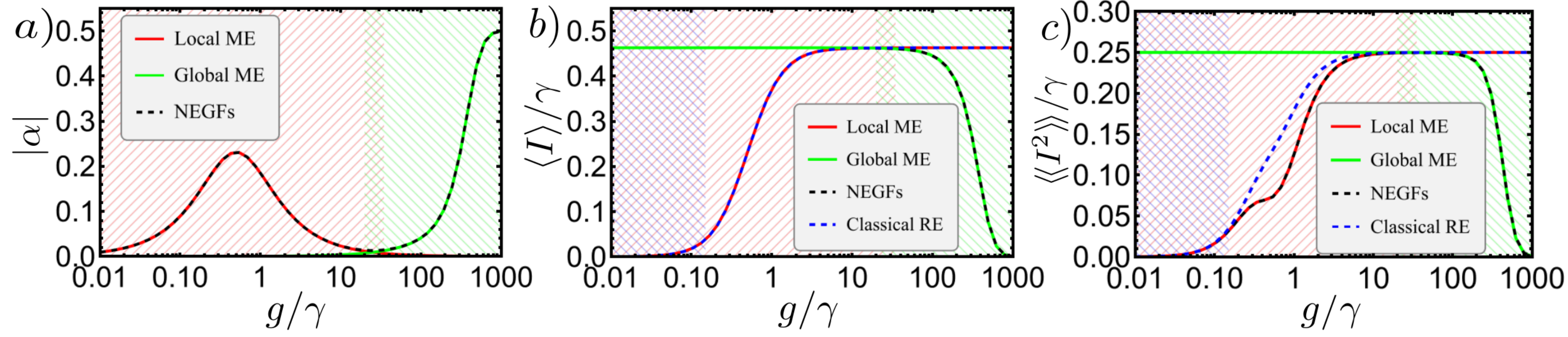}
        \captionsetup{justification = justified, singlelinecheck=false}
        \caption{Coherence, current, and fluctuations for noninteracting electrons. Results are shown for the local master equation (red), global master equation (green), classical model (blue dashed), and NEGFs (black dashed). (a) Coherence $|\alpha|$, (b) average current $\langle I \rangle$, and (c) current fluctuations $\langle\!\langle I^2 \rangle\!\rangle$, all as a function of $g/\gamma$. Parameters: $\gamma_L = \gamma_R = \gamma$, $T/\gamma = 100$, $\mu_L = eV/2$, $\mu_R = -eV/2$, $eV/T = 6.5$, $\epsilon = 0$, $U = 0$. Background shadings mark approximate regions where the local master equation (red), the global master equation (green), and the classical model (blue) are valid, respectively.}
        \label{fig: PanelA}
\end{figure*}

Finally, we consider the TKUR that combines the above bounds as \cite{Vo_2022}
\begin{equation}
\label{eq:TKUR}
     \mathcal{Q}_{TK} := \frac{\langle\!\langle I\rangle \! \rangle^2}{\langle\! \langle I^2 \rangle\! \rangle} \frac{4 \langle \mathcal{A} \rangle}{\langle \sigma \rangle^2} f\left( \frac{\langle \sigma \rangle}{2 \langle \mathcal{A} \rangle}\right)^2 \leq 1,
\end{equation}
where $f(x)$ is the inverse function of $x \tanh{x}$.
This inequality optimizes between the TUR and the KUR, and thus gives a stronger bound than both of them.

When investigating violations of uncertainty relations, we use the quantifier
\begin{equation}
\label{eq:Violation}
    \mathcal{V}_j = \text{max} \{\mathcal{Q}_j - 1, 0 \},
\end{equation}
where $j = T, K, TK$ correspond to TUR, KUR, and TKUR, respectively. We refer to this quantity as $j$ violation. If it is positive, the respective uncertainty relation is violated.
We note that numerically comparing the values of these quantifiers between different types of violations is not meaningful.


\section{Noninteracting electrons}
\label{sec:Non-interacting}
In this section, we provide our results for noninteracting electrons. In this case, there exist exact solutions obtained with NEGFs (cf. Appendix~\ref{App:NEGFs}), which serve as a benchmark. Before illustrating the manifestations of coherence, we consider the influence of the tunneling strength $g$ on coherence, the current, and its fluctuations. This allows for illustrating the coherent nature of the dynamics and to compare the different models. All analytical expressions provided in this section are derived with the local Lindblad master equation. For analytical results obtained with the global approach we refer the reader to Appendix~\ref{App: Global}.

\subsection{Coherent dynamics: Comparing the models}

We first discuss how the coherence in the system changes with the interdot tunnel coupling in the steady state.  We find
\begin{equation}
    \label{eq:Coherence Local}
|\alpha| = \frac{2g|n_L - n_R|\gamma_L \gamma_R}{(\gamma_L + \gamma_R)(4g^2 +  \gamma_L \gamma_R )}.
\end{equation}
Interestingly, this expression exhibits a peak at $g \simeq \gamma$ (with $\gamma_L = \gamma_R = \gamma$) [see Fig.~\ref{fig: PanelA}(a)]. This peak in coherence can be understood as follows: The coherent tunneling induces Rabi oscillations between the left and the right dots, which may result in a buildup of coherence. For $g\ll\gamma$, the decoherence induced by the bath suppresses any buildup of coherence akin to the Zeno effect. For $g\gg\gamma$, the Rabi oscillations become very fast, suppressing coherence by phase averaging. For intermediate $g$, coherence can build up, resulting in a peak. We note that the location of this peak does not depend on temperature or chemical potential (which only enter in the Fermi-Dirac distributions $n_\ell$).

While the peak in coherence is captured well by the local master equation, this approach breaks down when $g$ becomes of the order of $T$ [cf. Eq.~\eqref{eq: Condition local}]. In this regime, where the global master equation is justified, coherence grows again. As discussed above, this is a result of the entangled ground state being more populated than the excited state. In the large $g$ limit, the steady state reduces to the ground state, a pure singlet state with maximal coherence. As illustrated in Fig.~\ref{fig: PanelA}(a), we find excellent agreement between the exact NEGF solution and the Lindblad master equation solutions in their respective regime of validity.


Additionally, the current and its fluctuations provide insight into the effect of coherence on the dynamics. For the current, the local master equation yields
\begin{equation}
\label{eq:Current local}
    \langle I \rangle = \frac{4g^2 (n_L - n_R) \gamma_L \gamma_R (\gamma_L + \gamma_R)}{(\gamma_L + \gamma_R)^2(4g^2 + \gamma_L \gamma_R) }.
\end{equation}
The average current is illustrated in Fig.~\ref{fig: PanelA}(b). Starting from small $g$, its value grows as $g^2$ and saturates when the tunnel coupling $g$ becomes larger than the system-bath coupling $\gamma$, which becomes the bottleneck for transport. Similarly to the coherence, when $g$ is of the order of $T$ we observe a breakdown of the local approach. For larger values of $g$, where the global approach describes the dynamics well, the average current decreases, as the eigenenergies of the Hamiltonian leave the bias window, and transport of electrons is impeded. In addition, the plot includes the average current predicted by the classical model (cf. Appendix~\ref{App:Stochastic model}). It is in complete agreement with the local master equation, and as such is accurate in the regime of validity of the local approach. 

The current fluctuations are given by
\begin{equation} 
\label{eq:Variance local}
\begin{split}
     \langle\!\langle I^2 \rangle\!\rangle &= 
     \langle I \rangle \frac{n_L + n_R - 2n_L n_R}{n_L - n_R}\\
     &- 2\langle I \rangle^2 \left( \frac{1}{\gamma_L + \gamma_R} + \frac{\gamma_L + \gamma_R}{4g^2 + \gamma_L \gamma_R}\right),
\end{split}
\end{equation}
which is illustrated in Fig.~\ref{fig: PanelA}(c). Comparison with the NEGF solution shows that the local and global master equations reproduce the exact current fluctuations in their respective regime of validity. To the best of our knowledge, this is the first time that these regimes of validity have been benchmarked by quantities that do not only depend on the quantum state alone but instead depend on two-time correlation functions. As a function of $g/\gamma$, the fluctuations display analogous features to the average. However, when $g$ is of the same order as $\gamma$, the fluctuations are suppressed below the value of the classical model (see also Ref.~\cite{Kiesslich_2006} for a similar observation). The regime characterized by the peak in coherence thus coincides with the regime where the fluctuations of the current are suppressed relative to the classical model. These features underlie the manifestations of coherence that we focus on below.


\begin{figure*}[t]
    \centering
        \includegraphics[width=\textwidth]{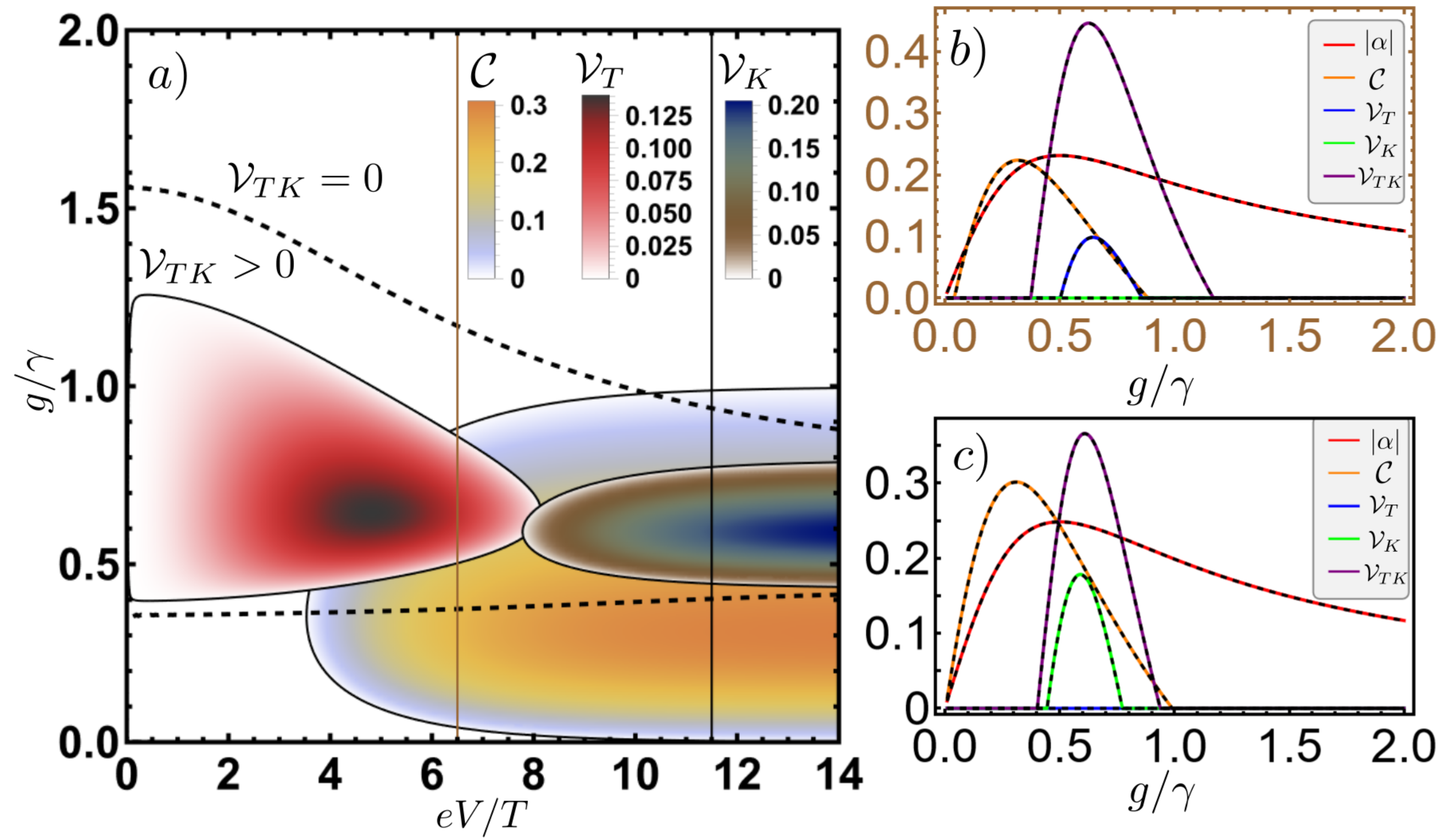}
    \caption{Comparison of coherence $|\alpha|$, concurrence $\mathcal{C}$, TUR violation $\mathcal{V}_T$, KUR violation $\mathcal{V}_K$, and TKUR Violation $\mathcal{V}_{TK}$ for noninteracting electrons. (a) Manifestations of coherence as a function of $g/\gamma$ and $eV/T$. Results are obtained with the local master equation. Parameters: $\gamma_L = \gamma_R = \gamma$, $T/\gamma = 100$, $\mu_L = eV/2$, $\mu_R = -eV/2$, $\epsilon = 0$, $U = 0$. For clarity we only show the region, enveloped by the black dashed lines, where $\mathcal{V}_{TK} > 0$. [(b), (c)] Cross sections of panel (a) (gold and black vertical lines) corresponding to $eV/T = 6.5$ and $eV/T = 11.5$, respectively.}
        \label{fig: TUR Concurrence}
\end{figure*}



\subsection{Manifestations of coherence}
\label{Sec: Manifestations of coherence}

The expression for the concurrence [cf. Eq.~\eqref{eq:Concurrence}] is given by
\begin{align}
    \mathcal{C} =  \max \Bigg\{0,~&2 |\alpha| - \frac{2 \gamma_L \gamma_R}{4g^2 + \gamma_L \gamma_R}   \sqrt{\left[n_L n_R   + \frac{4g^2}{\gamma_L \gamma_R}\bar{n}^2\right]}
    \nonumber \\[0.2cm]
    &\times \sqrt{\left[(1-n_L)(1-n_R)  +\frac{4g^2}{\gamma_L \gamma_R}(1-\bar{n})^2\right]}\Bigg\},
\label{eq:Concurrence Local}    
\end{align}
where 
\begin{equation}
    \bar{n} = \frac{n_L \gamma_L + n_R \gamma_R}{\gamma_L + \gamma_R},
\end{equation}
and $|\alpha|$ is given in Eq.~\eqref{eq:Coherence Local}. Maximal concurrence occurs for infinite bias, $(\epsilon - \mu_L)/T_L \to -\infty$ and $(\epsilon - \mu_R)/T_R \to \infty$; i.e., $eV\rightarrow\infty$. In this regime the maximal value is $\mathcal{C} = (\sqrt{5} - 1)/4 \approx 0.31$ and can be reached by setting $g/\gamma = (\sqrt{5} - 1)/4$. 
The TUR violation, on the other hand, is computed from Eq.~\eqref{eq:Violation} together with Eq.~\eqref{eq:TUR2}.
It becomes saturated at $g/\gamma = \sqrt{15}/6$~\cite{Ptaszynski_2018} irrespective of the applied bias, similarly as it was found for the coherence. The corresponding maximum is $\mathcal{V}_T \approx 0.141$, which occurs at $eV/T \approx 4.8$. 
Similarly, we find KUR violations, i.e., $\mathcal{V}_K>0$. Its optimal value $\mathcal{V}_K \approx 0.216$ occurs when $(\epsilon - \mu_L)/T_L \to \infty$ and $(\epsilon - \mu_R)/T_R \to -\infty$, which is the same regime saturating concurrence, and when $g/\gamma \approx 0.59$.

All three manifestations are presented in Fig.~\ref{fig: TUR Concurrence} as a function of $g$ and $eV$. 
The range of values of $g$ coincides with the window of the tunnel couplings that exhibit both a peak of coherence and reduction of current fluctuations. This is directly demonstrated in Figs.~~\ref{fig: TUR Concurrence}(b) and ~\ref{fig: TUR Concurrence}(c), where we compare $|\alpha|$ with $\mathcal{C}$, $\mathcal{V}_T$, $\mathcal{V}_K$, and $\mathcal{V}_{TK}$ on two cross sections of Fig.~\ref{fig: TUR Concurrence}(a). Both plots show excellent agreement between the local master equation (solid color lines) and NEGFs (corresponding dashed black lines). The regime where $g$ and $\gamma$ are comparable thus features both entanglement, a static manifestation of coherence describing the steady state, as well as TUR and KUR violations, dynamical manifestations of coherence which arise from the nonclassical dynamics of the system. Crucially, we see how these two manifestations of coherence do not always appear together. As discussed above, in the regime of strong $g$, entanglement is found. However, the dynamics of the system are well described by the global master equation, a classical rate equation involving the entangled eigenstates of the Hamiltonian. Consequently, we do not find TUR and KUR violations. 

There are also regimes where entanglement and TUR violations are found to partially overlap, such that a crossover can be achieved by increasing the voltage bias. Specifically, the TUR is only violated for small voltage bias, whereas entanglement requires large voltage bias. Increasing $eV$ gives rise to increasing entropy production, which causes the TUR to become less tight. Entanglement grows with increasing current, which itself increases with the voltage. For large voltages, we also observe KUR violations due to the suppression of current fluctuations. In agreement with previous findings $\mathcal{V}_K \approx 0.216$, TUR violations are observed close to equilibrium while KUR violations appear far from equilibrium. As the TKUR is tighter than both the TUR and the KUR, its violations encompass both the TUR as well as the KUR violations. 

We end this section by considering the usefulness of the entangled quantum state to achieve nonlocality \cite{Brunner_2014} and to perform quantum teleportation \cite{Bennet_1993} (see Appendix~\ref{app:opnonclass}).  This has been done previously in a similar system \cite{Brask_2022}. 
For noninteracting electrons, we find a teleportation fidelity of $f = (7 + \sqrt{5})/12 \approx 0.77$ (cf. Appendix~\ref{App:Teleportation}), which is above the classical limit of $2/3$. As expected, this corresponds to the value found in Ref. \cite{Brask_2022} in the noninteracting case. 
Concerning nonlocality, in our system with noninteracting electrons we do not find any violation of the Clauser-Horne-Shimony-Holt (CHSH) inequality \cite{Clauser} (cf. Appendix~\ref{App:Nonlocality}), in agreement with the findings of Ref.~\cite{Brask_2022}.
As discussed in Sec.~\ref{Sec:Nonlocality}, inclusion of Coulomb interactions allows for a higher teleportation fidelity and even CHSH violations, implying Bell nonlocality. This is in contrast to the results of Ref.~\cite{Brask_2022}, where no chemical potential was present and population inversion was obtained using negative temperatures.

\section{Interacting electrons: Enhanced coherence, stronger violations, and nonlocal states}
\label{sec:Interacting}

\begin{figure*}[t]
    \centering
        \includegraphics[width=1\textwidth]{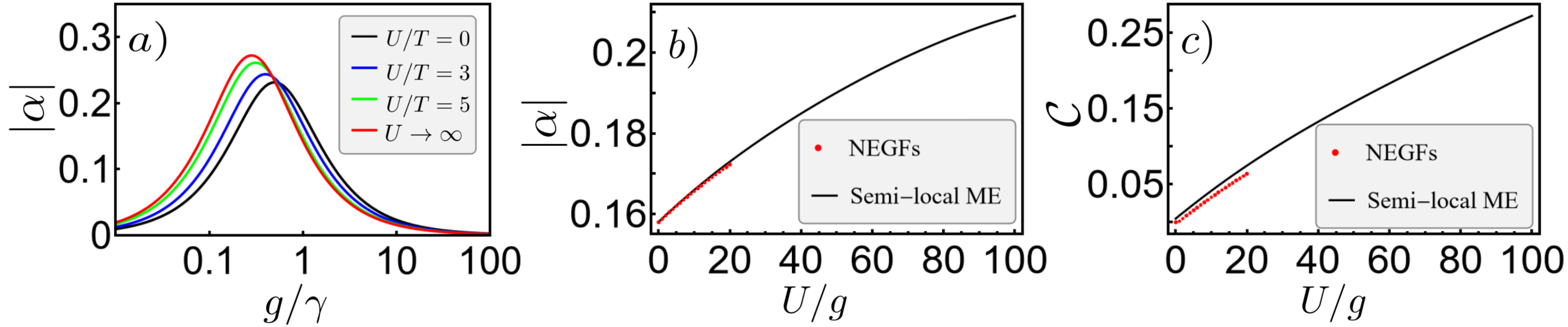}
        \caption{The effect of Coulomb interactions on coherence and concurrence. (a) Coherence $|\alpha|$ as a function of $g/\gamma$ for different strength of Coulomb interaction: $U/T = 0$ (black), $U/T = 3$ (blue), $U/T = 5$ (green), and $U \to \infty$ (red). The plot was produced with the results of the semilocal master equation. Parameters: $T/\gamma = 100$, $eV/T = 6.5$, $\epsilon = 0$. [(b), (c)] Coherence $|\alpha|$ and concurrence $\mathcal{C}$ as a function of $U/g$, respectively, obtained with the semilocal master equation (black) and NEGFs (red). Parameters: $g/\gamma = 0.3$, $T/\gamma = 10$, $eV/T = 3.6$, and $\epsilon = 0$.}
       \label{fig:U - coh and conc}
\end{figure*}


Coulomb interactions between the electrons change the dynamics of the double QD system by incorporating an additional energetic cost for occupying both QDs. In this section we show how the presence of interactions affects the manifestations of coherence.

\subsection{Methods}
We investigate the dynamics of the system using methods analogous to the noninteracting case, adjusted to take into account the effect of Coulomb interactions between electrons.

\subsubsection{Master equation}
In the previous section we have identified the regime of interest to be where the peak of coherence appears together with all the manifestations of coherence. Since the same holds for interacting electrons, we restrict our attention to the regime where tunnel couplings $g$ are comparable to $\gamma_\ell$. A Lindblad master equation suitable for treating this regime in the presence of Coulomb interactions (termed semilocal) has recently been microscopically derived \cite{Potts_2021}.
The corresponding superoperators read
    \begin{equation}
    \label{eq:Lindblad semilocal}
    \begin{split}
        \mathcal{L}_\ell &= \gamma_\ell \left[n_\ell \mathcal{D}[(1-\hat{c}_{\bar{\ell}}^{\dagger}\hat{c}_{\bar{\ell}}) \hat{c}_\ell^{\dagger}] + (1-n_\ell) \mathcal{D}[(1-\hat{c}_{\bar{\ell}}^{\dagger}\hat{c}_{\bar{\ell}}) \hat{c}_\ell] \right] \\
        & + \gamma_\ell \left[n_\ell^U \mathcal{D}[\hat{c}_{\bar{\ell}}^{\dagger}\hat{c}_{\bar{\ell}} \hat{c}_\ell^{\dagger}]  + (1-n_\ell^U) \mathcal{D}[\hat{c}_{\bar{\ell}}^{\dagger}\hat{c}_{\bar{\ell}} \hat{c}_\ell] \right],
    \end{split}
\end{equation}
where
\begin{equation}
    n_\ell := n_\ell(\epsilon),
\end{equation}
\begin{equation}
    n_\ell^U := n_\ell(\epsilon + U),
\end{equation}
and $\bar{\ell} \neq \ell$.
It is valid in the regime
\begin{equation}
\label{eq: Condition semi-local}
    g \ll \max\{T_\ell, | \epsilon + U - \mu_\ell|\} \qquad \text{(semilocal ME)}.
\end{equation}
The key difference between the semilocal master equation and the local one, which we used for the noninteracting case, is that now the jump rates from each QD to the adjacent reservoir depend on the occupation of the other QD. Importantly, the results from the semilocal master equation reduce to the results from the local master equation in the limit $U\rightarrow0$ \cite{Potts_2021}.

In order to compute the average current and its fluctuations we proceed analogously to the case of noninteracting particles. The average current in the steady state is given by Eq.~\eqref{eq: Current SS},
and the fluctuations are obtained with full counting statistics (Appendix~\ref{App:FCS}). 

\subsubsection{Classical model}

As for noninteracting electrons, we derived a classical model of the form of Eq.~\eqref{eq:SM} using the perturbative approach described in Appendix~\ref{App:Stochastic model}. 
The matrix elements describing jumps of electrons between the reservoirs and the system read $W_{L0} = \gamma_L n_L$, $W_{0L} = \gamma_L (1-n_L)$, $W_{DR} = \gamma_L n_L^U$, $W_{RD} = \gamma_L (1-n_L^U)$, and similarly for $L \leftrightarrow R$. These are the same rates that appear in the semilocal master equation [cf. Eq.~\eqref{eq:Lindblad semilocal}]. The interdot tunnel rate is given by
\begin{equation}
    W_{LR}  = \frac{4g^2}{\gamma_L(1- n_L + n_L^U) + \gamma_R(1- n_R + n_R^U)},
\end{equation}
and $W_{RL} = W_{LR}$.

\subsubsection{NEGFs}
\label{Sec:NEGFU main}
The effect of Coulomb interactions in the NEGF can be introduced via
a many-body self-energy. This is an in-principle exact procedure. In practice, one must, however resort to approximate self-energies, calculated, e.g., via many-body approximations. Here, we consider a rather simple approximation, namely the second Born perturbative scheme~\cite{Kadanoff1962,Baym1962,Keldysh1965} (for details we refer the reader to Appendix~\ref{App:NEGF2B}). The range of $U/g$ where the second Born approximation
gives a reliable description of electronic correlations is limited and much smaller
than the range we consider with the Lindblad master equation in the remainder of the paper.
Nonetheless, it may still be useful to compare
the two methods in a restricted domain where both approaches produce meaningful results.

We thus notice that in this limited $U/g$ range the two treatments give results that are
in good mutual agreement for the coherences, while discernible discrepancies occur for the concurrence,
due to well-known possible shortcomings of the second Born approximation when determining double occupancies
(see Appendix~\ref{App:NEGF2B}).

\begin{figure*}[t]
    \centering
        \includegraphics[width=1\textwidth]{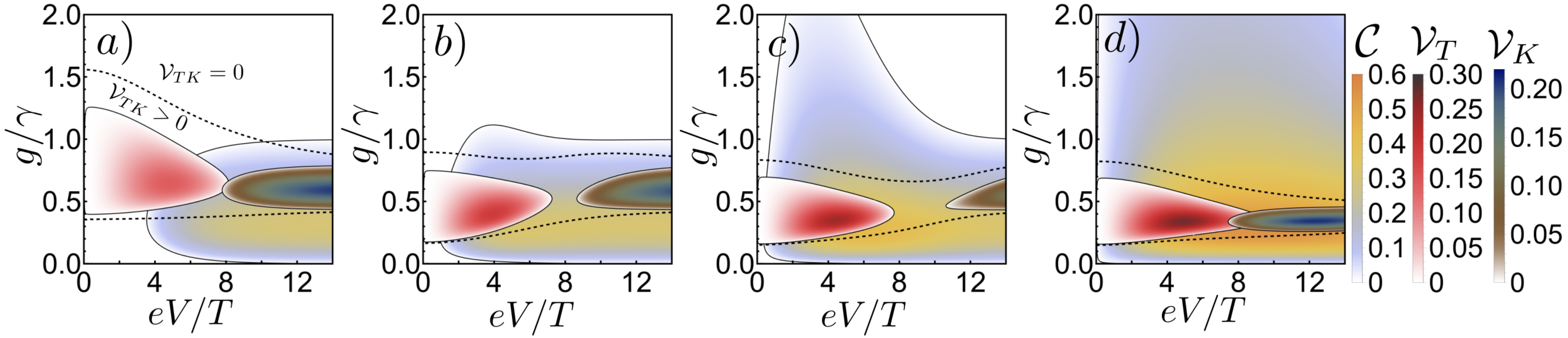}
        \caption{Evolution of concurrence $\mathcal{C}$, TUR violation $\mathcal{V}_T$, KUR violation $\mathcal{V}_K$, and TKUR violation $\mathcal{V}_{TK}$ for different strengths of Coulomb interaction: (a) $U/T = 0$, (b) $U/T = 3$, (c) $U/T = 5$, and (d) $U/T = 10$. Plots as a function of $g/\gamma$ and $eV/T$ were obtained with the local master equation. Parameters: $\gamma_L = \gamma_R = \gamma$, $T/\gamma = 100$, $\mu_L = eV/2$, $\mu_R = -eV/2$, $\epsilon = 0$. Dashed black lines envelop the regions of TKUR violations.}
       \label{fig: TUR Concurrence U}
\end{figure*}

\subsection{Enhanced manifestations of coherence}

Since Coulomb interactions introduce an energetic cost of having both QDs simultaneously filled, the probability of double occupation, $p_D$, is suppressed and vanishes in the limit $U \to \infty$.
This reduced occupation of the doubly filled subspace generally has a beneficial effect on the amount of coherence, as it increases the probability for the subspace with a single electron on the dot. However, as explained below, Coulomb interactions do not always result in an increased amount of coherence.
The expression for coherence reads
\begin{equation}
\begin{split}
    |\alpha| =& \frac{2g \gamma_L \gamma_R}{\gamma_L + \gamma_R} \\
    & \times \frac{  n_L(1-n_R^U) - n_R(1-n_L^U)}{ 4g^2(1 + \bar{n} - \bar{n}_U ) + (1-\Delta_L \Delta_R)\gamma_L \gamma_R(1 - \bar{n} + \bar{n}_U)},
\end{split}
\end{equation}
where we have introduced
\begin{equation}
    \bar{n}_U = \frac{\gamma_L n_L^U + \gamma_R n_R^U}{\gamma_L + \gamma_R},
\end{equation}
and $\Delta_l = n_l - n_l^U$.
In the limit $U \to \infty$ the coherence reduces to
\begin{equation}
    |\alpha| = \frac{2g \gamma_L \gamma_R \left( n_L - n_R\right)}{(\gamma_L + \gamma_R)\left[ 4g^2(1 + \bar{n} ) + (1-n_L n_R)\gamma_L \gamma_R(1 - \bar{n})\right]}.
\end{equation}
As illustrated in Fig.~\ref{fig:U - coh and conc}(a), increasing $U$ results in a larger peak in coherence that is shifted to smaller values of $g/\gamma$. Because of this shift, there are values of $g/\gamma$ where coherence decreases when increasing $U$.

From the definition of concurrence, given by Eq.~\eqref{eq:Concurrence}, we anticipate that Coulomb interactions increase the amount of entanglement in the system for two reasons: First, because $p_D$ is smaller, and second, because this may increase the off-diagonal element $\alpha$. This is illustrated in Figs.~\ref{fig:U - coh and conc}(b) and~\ref{fig:U - coh and conc}(c), where we survey the influence of $U/g$ on the coherence and concurrence, respectively. In Fig.~\ref{fig:U - coh and conc}(b) coherence shows moderate growth with the strength of the Coulomb interaction. For comparison, we feature the results of the semilocal master equation (black) and of NEGFs (red) on the restricted interval, demonstrating a good agreement between the two methods. A similar plot showing the concurrence is presented in Fig.~\ref{fig:U - coh and conc}(c). For concurrence we observe a strong impact of Coulomb interactions, which indicates that the growth of concurrence comes predominantly from the suppression of $p_D$. We note that the agreement between the semilocal master equation and NEGFs is better for coherence than for concurrence. We attribute this to the fact that the concurrence relies not only on single particle quantities but requires the computation of $p_D$. As discussed in Appendix~\ref{App:NEGF2B}, this quantity may be more sensitive to the employed approximations.

In addition to entanglement, Coulomb interactions impact the dynamical manifestations, TUR and KUR violations, which is illustrated in Fig.~\ref{fig: TUR Concurrence U}. In the series of panels we show how $\mathcal{C}$, $\mathcal{V}_T$, and $\mathcal{V}_K$ evolve as $U/T$ increases. Concurrence and TUR violations are considerably enhanced by interactions, while the enhancement of KUR violations is less prominent. While the range of $g/\gamma$ where entanglement is present is extended, the same is not true for the remaining manifestations. This is a consequence of the fact that larger $\mathcal{C}$ is caused predominantly by the reduction of $p_D$ as opposed to an increase in coherence. 
In addition, the values of $g/\gamma$ that maximize $\mathcal{V}_T$ and $\mathcal{V}_K$ become smaller and align with the optimal value for $\mathcal{C}$. This is similar to the observed shift in the peak of coherence.

\subsection{Nonlocality and entanglement for an interacting double uqantum dot}
\label{Sec:Nonlocality}

The main drawback of looking at the entanglement solely through the prism of any entanglement measure, such as concurrence, is that it does not provide a definite measure of the usefulness to achieve nonlocality~\cite{Brunner_2014} or perform quantum information tasks, such as quantum teleportation~\cite{Bennet_1993}. The problem of generating operational nonclassicality in the class of systems akin to the double quantum dot was systematically analyzed by Bohr Brask \textit{et al.}~\cite{Brask_2022}, where no nonlocality was found. In their approach, however, the system was brought out of equilibrium via a temperature gradient alone but allowing for negative temperatures to describe population inversion. Here we show that, by considering a bias voltage, Coulomb interactions may result in nonlocality.

The optimal regime to maximize concurrence and observe nonlocality is provided by the following conditions: First, jumps from the right reservoir into the system are completely suppressed, i.e., $n_R = n_R^U = 0$. Second, jumps from the left reservoir are suppressed only when there is already an electron in the system, i.e., $n_L^U = 0$ and $n_L = 1$. This is achieved in the limits $U \to \infty$ and $eV \to \infty$, with $U/eV \to \infty$.



Under these conditions, the probability of the double occupation $p_D$ vanishes and the coherence reduces to
\begin{equation}
\label{Eq:CohUinf}
    |\alpha| = \frac{2 g \gamma_L \gamma_R}{\gamma_L \gamma_R^2 + 4g^2 (2 \gamma_L + \gamma_R)}.
\end{equation}
The expression for concurrence simplifies to
\begin{equation}
\label{Eq:p0Uinf}
    \mathcal{C} = \frac{4g \gamma_L \gamma_R}{\gamma_L \gamma_R^2  + 4g^2 (2 \gamma_L + \gamma_R)}.
\end{equation}
Upon examining $\frac{\partial{\mathcal{C}}}{{\partial{g}}}$,
we find the maximal value $\mathcal{C} = \sqrt{2}/2 \approx 0.71$ when $\gamma_R/g = 2 \sqrt{2}$ and $\gamma_L/g \to \infty$. This is significantly larger than $(\sqrt{5}-1)/4 \approx 0.31$, which we found to be the maximum in the noninteracting case. If the couplings to the baths are assumed to be equal, the largest concurrence occurs at $\gamma/g = 2 \sqrt{3}$, resulting in
$\mathcal{C} = \sqrt{3}/3 \approx 0.58$. We note that in the absence of a chemical potential (even allowing for negative temperatures), Coulomb interactions do not alter the maximal value for the concurrence \cite{Brask_2022}.

Motivated by the substantial increase in concurrence due to Coulomb interactions, we revisit the question of Bell nonlocality (cf. Appendix~\ref{App:Nonlocality}), focusing on the CHSH scenario \cite{Clauser}, which was also considered in Ref.~\cite{Brask_2022}. In this scenario, if the statistics of the outcomes of measurements on a quantum state admit a local hidden-variable model, then the CHSH quantity obeys $\text{CHSH} \leq 2$~\cite{Brunner_2014}. Any violation of this bound is a signature of nonlocality, and the maximal $\text{CHSH} = 2 \sqrt{2}$ can be achieved with maximally entangled Bell states. For the double quantum dot, in the regime discussed above, we find the maximal value $\text{CHSH} = 2 \sqrt{3/2}$, which is achieved for the same set of parameters $\gamma_L, \gamma_R$, and $g$ that saturate concurrence. When $\gamma_R/g = 2\sqrt{2}$, CHSH displays nonlocality for $\gamma_L/g > 5.66$.

For the optimal fidelity of quantum teleportation~\cite{Horodecki_1999} we find $f = (4 + \sqrt{2})/6 \approx 0.9$, which outperforms the system without Coulomb interactions, where $f = (7 + \sqrt{5})/12 \approx 0.77$, which coincides with the maximum found in Ref.~\cite{Brask_2022}. The details of the calculations are outlined in Appendix~\ref{App:Teleportation}. 

\section{More general models}
\label{sec:Extended}

\begin{figure*}[t]
    \centering
        \includegraphics[width=1\textwidth]{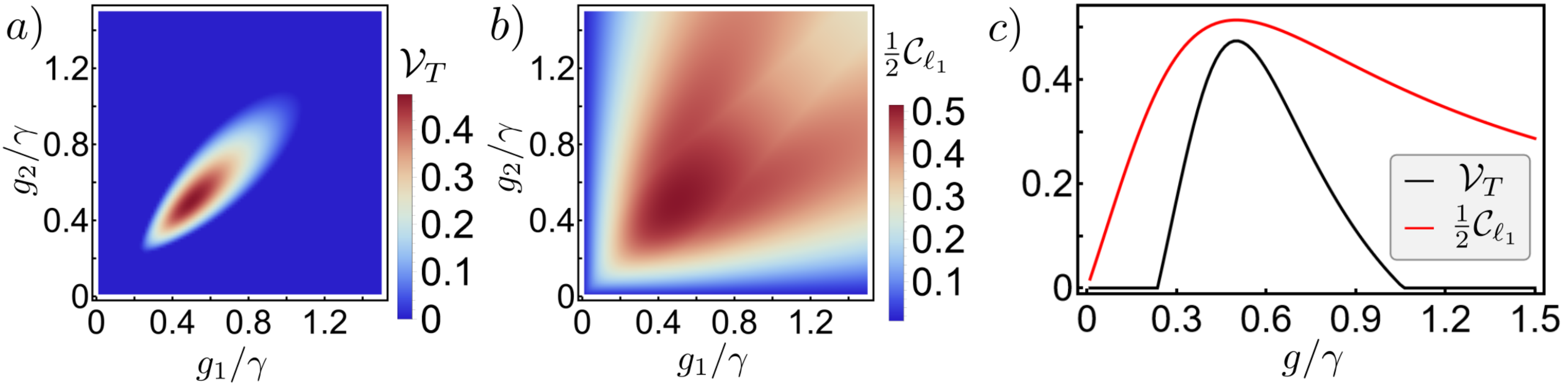}
        \caption{TUR violations and coherence in the chain of three QDs: (a) TUR violation, $\mathcal{V}_T$, and (b) one-half of the $\ell_1$-norm of coherence, $\frac{1}{2}\mathcal{C}_{\ell_1}$, as a function of two coherent couplings $g_1/\gamma$ and $g_2/\gamma$; c) diagonal cross-sections corresponding to $g_1 = g_2 = g$. Parameters: $eV/T = 5$, $T/\gamma =100$, $\epsilon = 0$.}
       \label{fig: Extended}
\end{figure*}

Up to this point we have been concerned with the minimal model consisting of only two sites, which allows to clearly identify the regimes of nonclassical behavior. The fact that similar conclusions hold for both interacting and noninteracting electrons implies that our results may serve to anticipate the behavior of more general mesoscopic systems. To illustrate this, we investigate TUR violations in a chain of three QDs, where electrons in the system may coherently traverse between the neighboring QDs.
The corresponding Hamiltonian is given by
\begin{equation}
    \hat{H} = \sum_{\ell = L, M, R} \epsilon \hat{c}_\ell^\dagger c_\ell + g_1 (\hat{c}_L^\dagger c_M + \hat{c}_M^\dagger c_L) + g_2 (\hat{c}_M^\dagger c_R + \hat{c}_R^\dagger c_M),
\end{equation}
where the subscript $M$ refers to the middle QD, and $g_1$ ($g_2$) denotes the coherent coupling between the left and the middle (the right and the middle) QDs. We investigate the chain with the NEGFs using analogous methods as those presented in Appendix~\ref{App:NEGFs}.
As anticipated from our simple model, we find nonclassical behavior where both coherent couplings, $g_1$ and $g_2$, are of the order of the coupling to the baths $\gamma$. This is illustrated in Figs.~\ref{fig: Extended}(a) and~~\ref{fig: Extended}(b), which show TUR violation $\mathcal{V}_T$ and the amount of coherence in the system, respectively. As a quantifier of the amount of coherence we use one-half of the $\ell_1$-norm of coherence, here denoted by $C_{\ell_1}$, which reduces to $|\alpha|$ in the two-site model. The peaks of $\mathcal{V}_T$ or $C_{\ell_1}$ are visualized in Fig.~\ref{fig: Extended}(c), where we show the diagonal cross-sections with equal coherent couplings, $g_1 = g_2 = g$, of Figs.~\ref{fig: Extended}(a) (black) and~\ref{fig: Extended}(b) (red), respectively. These graphs further demonstrate that the TUR violation and coherence are maximized for $g_1 = g_2$ both being comparable to $\gamma$. Moreover, in the chain of three QDs we observe stronger TUR violation than in a double quantum dot (DQD) operating under the same set of parameters (see Fig.~\ref{fig: TUR Concurrence}). This indicates that nonclassical behavior is more easily accessible in longer chains. The fact that our conclusions, drawn from the investigation of the DQD, hold in the chain of three QDs provides evidence for the applicability of these results in more general systems. We note that these results are also consistent with Ref.~\cite{Saryal_2019, Schaller_2021}, where TUR violations in longer chains of QDs were investigated.

\section{Conclusions and outlook}
\label{sec:Conclusions}

In this paper we investigated the nonclassical behavior of a mesoscopic open quantum system by analyzing two complementary manifestations of coherence: Entanglement, a static manifestation, and violations of thermokinetic uncertainty relations, dynamical manifestations of coherence. We considered a double quantum dot weakly coupled to fermionic reservoirs, where the transport of electrons between the dots can be modeled either by coherent tunneling, or by stochastic jumps. 


We identified the regime where the system exhibits truly nonclassical behavior: On the one hand, coherence exhibits a peak, which may result in an entangled state and, in the presence of interactions, even nonlocality. On the other hand, coherent tunneling suppresses the fluctuations of the current below the classical model, allowing for violations of the TUR and the KUR. This behavior occurs in the regime where the coherent tunnel-coupling is of the order of the coupling to the baths, which is captured by the local master equation. This is different from the regime of large tunnel couplings, which is captured by the global master equation, where the state of the system may be strongly entangled, but the dynamics of the system can be expressed through a classical rate equation, and no TUR or KUR violations are present.


Our systematic characterization of the electron transport across the system performed with different master equations and NEGFs illustrates the need to go beyond the steady state in order to fully capture the nonclassical behavior in mesoscopic transport. While the steady state captures the entanglement, the violations of thermokinetic uncertainty relations cannot be captured by the steady state alone. Those different manifestations of coherence are not equivalent and do not need to coincide. We note that in the regime of nonclassical behavior, not only the quantum state of the DQD but also the current fluctuations, which depend on two-time correlations, are well captured by the local master equation. While our results are obtained for a specific model, we believe that these conclusions are relevant for a broad range of open quantum systems and provide guiding principles for the design of out-of-equilibrium devices that exhibit nonclassical transport. This is evidenced by the fact that our predictions about the nonclassical behavior, drawn from the noninteracting two-site model, also hold for interacting electrons and in the chain of three QDs. 

Our results may be probed experimentally in a number of different platforms, including electronic QDs, which can be implemented, e.g., using two-dimensional electron gases~\cite{exp_Peterson, exp_Frey} or nanowires~\cite{exp_BarkerNW, exp_Nilsson}, as well as superconducting qubits~\cite{exp_Aamir_2022, flux_Izmalkov, flux_Majer}. While measurements of the electric current and its fluctuations across a DQD is a standard experimental technique~\cite{exp_Okazaki}, accessing the quantum state of the DQD is more challenging. Alternatively, the density matrix and, therefore, the entanglement may be probed in an implementation of our model using two coupled superconducting qubits \cite{exp_Aamir_2022}.

\begin{acknowledgments}
We acknowledge contributions by Meeri Mölsä during her bachelor thesis in the early stages of this work.
P.P.P. and K.P. acknowledge funding from the Swiss National Science Foundation (Eccellenza Professorial Fellowship PCEFP2\_194268). P.S. and C.V. acknowledge support from the Swedish 
Research Council, Grants No. 2018-03921 and No. 2017-03945, respectively.
\end{acknowledgments}

\appendix

\section{Noninteracting electrons}
\label{App:Non-interacting}
In the main text of the paper we restrict our attention to the case with the same onsite energies of both QDs. In this appendix, however, we provide a derivation of the results in the general case of noninteracting electrons. The Hamiltonian operator of the system is given by 
\begin{equation}
\label{eq:He}
\hat{H} = \sum_{\ell = L, R} \epsilon_\ell \hat{c}_\ell^{\dagger} \hat{c}_\ell  + g \left( \hat{c}_L^{\dagger} \hat{c}_R + \hat{c}_R^{\dagger}\hat{c}_L \right),
\end{equation}
where $\epsilon_\ell$ is the on-site energy of the QD $\ell$. In the noninteracting case, the coefficients of density matrix \eqref{eq:Density Matrix} can be expressed as
\begin{equation}
\label{eq:Coefficients}
    \begin{split}
         &p_0 = \left(1- \langle \hat{c}_L^{\dagger} \hat{c}_L \rangle\right) \left(1 - \langle \hat{c}_R^{\dagger} \hat{c}_R \rangle \right) -  \langle \hat{c}_L^{\dagger}  \hat{c}_R \rangle \langle \hat{c}_R^{\dagger} \hat{c}_L \rangle,\\
        &p_L = \langle \hat{c}_L^{\dagger} \hat{c}_L \rangle \left(1 - \langle \hat{c}_R^{\dagger} \hat{c}_R \rangle \right) + \langle \hat{c}_L^{\dagger}  \hat{c}_R \rangle \langle \hat{c}_R^{\dagger} \hat{c}_L \rangle,\\
        &p_R = \langle \hat{c}_R^{\dagger} \hat{c}_R \rangle \left(1- \langle  \hat{c}_L^{\dagger} \hat{c}_L \rangle \right) + \langle \hat{c}_L^{\dagger}  \hat{c}_R \rangle \langle \hat{c}_R^{\dagger} \hat{c}_L \rangle,\\
        &p_D =  \langle \hat{c}_L^{\dagger}  \hat{c}_L \rangle \langle \hat{c}_R^{\dagger} \hat{c}_R \rangle -  \langle \hat{c}_L^{\dagger}  \hat{c}_R \rangle \langle \hat{c}_R^{\dagger} \hat{c}_L \rangle,\\
        &\alpha = \langle \hat{c}_R^{\dagger} \hat{c}_L \rangle,
    \end{split}
\end{equation}
where we have used a corollary of Wick's theorem~\cite{Molinari},
\begin{equation}
    \langle \hat{c}_L^{\dagger} \hat{c}_L \hat{c}_R^{\dagger} \hat{c}_R \rangle = \langle \hat{c}_L^{\dagger} \hat{c}_L \rangle  \langle \hat{c}_R^{\dagger} \hat{c}_R \rangle - \langle \hat{c}_L^{\dagger} \hat{c}_R \rangle \langle \hat{c}_R^{\dagger} \hat{c}_L \rangle.
\end{equation}

\subsection{Local master equation}
\label{App: Local}
In the local master equation satisfying thermodynamic consistency the Lindblad superoperators appearing in Eq.~\eqref{eq:LME} are given by \cite{Potts_2021}
\begin{equation}
\label{eq:Locale}
    \mathcal{L}_\ell = n_\ell \gamma_\ell \mathcal{D}[\hat{c}^{\dagger}_\ell] + (1-n_\ell) \gamma_\ell \mathcal{D}[\hat{c}_\ell],
\end{equation}
where now $n_\ell := n_\ell(\bar{\epsilon})$, and $\bar{\epsilon} = (\epsilon_L + \epsilon_R)/2$.
In order to find the density matrix in the steady state, we consider the vector $\Vec{v} = [\langle \hat{c}_L^{\dagger} \hat{c}_L \rangle, \langle \hat{c}_R^{\dagger} \hat{c}_R \rangle, \langle \hat{c}_L^{\dagger} \hat{c}_R \rangle, \langle \hat{c}_R^{\dagger} \hat{c}_L \rangle]^T$. Its equation of motion reads
\begin{equation}
    \frac{d}{dt}\Vec{v} = A\Vec{v} + \Vec{b},
\end{equation}
where
\begin{equation}
\label{eq:Local motion}
    A = \begin{pmatrix}
    -\gamma_L & 0 & -ig & ig\\
    0 & -\gamma_R & ig & -ig\\
    -ig & ig & 2i\delta - \frac{\gamma_L + \gamma_R}{2} & 0\\
    ig & -ig & 0 & -2i\delta - \frac{\gamma_L + \gamma_R}{2}
    \end{pmatrix},
\end{equation}
$\Vec{b} = [\gamma_L n_L, \gamma_R n_R, 0, 0]^T$, and $\delta = (\epsilon_L - \epsilon_R)/2$. The steady-state solution $-A^{-1}\Vec{b}$ then yields
\begin{widetext}
\begin{equation}
\label{eq:Solution local}
    \begin{split}
        &\langle \hat{c}_L^{\dagger} \hat{c}_L \rangle = \frac{4g^2 (\gamma_L + \gamma_R)(n_L \gamma_L + n_R \gamma_R) + n_L \gamma_L \gamma_R\left((\gamma_L + \gamma_R)^2 + 16 \delta^2\right) }{(\gamma_L + \gamma_R)^2(\gamma_L \gamma_R + 4g^2) + 16 \gamma_L \gamma_R \delta^2},\\
        &\langle \hat{c}_R^{\dagger} \hat{c}_R \rangle = \frac{4g^2 (\gamma_L + \gamma_R)(n_L \gamma_L + n_R \gamma_R) + n_R \gamma_L \gamma_R\left((\gamma_L + \gamma_R)^2 + 16 \delta^2\right) }{(\gamma_L + \gamma_R)^2(\gamma_L \gamma_R + 4g^2) + 16 \gamma_L \gamma_R \delta^2},\\
        &\langle \hat{c}_L^{\dagger} \hat{c}_R \rangle = \frac{-2ig(n_L - n_R)\gamma_L \gamma_R(\gamma_L + \gamma_R + 4i\delta)}{(\gamma_L + \gamma_R)^2(\gamma_L \gamma_R + 4g^2) + 16 \gamma_L \gamma_R \delta^2},\\
        &\langle \hat{c}_R^{\dagger} \hat{c}_L \rangle = \frac{2ig(n_L - n_R)\gamma_L \gamma_R(\gamma_L + \gamma_R - 4i\delta)}{(\gamma_L + \gamma_R)^2(\gamma_L \gamma_R + 4g^2) + 16 \gamma_L \gamma_R \delta^2}.
    \end{split}
\end{equation}
Using Eq.~\eqref{eq:Coefficients} we obtain
\begin{equation}
\label{eq:Coefficients local}
    \begin{split}
        &p_0 
        = \frac{4g^2 (1-\bar{n})^2  + (1-n_L) (1-n_R) \gamma_L \gamma_R\left(1 +  \xi^2\right) }{4g^2 +  \gamma_L \gamma_R(1+ \xi^2)},\\
        &p_D 
        = \frac{4g^2 \bar{n}^2  + n_L n_R \gamma_L \gamma_R\left(1 +  \xi^2\right) }{ 4g^2 +  \gamma_L \gamma_R (1+\xi^2)},\\
        &p_L 
        = \frac{4g^2 \bar{n}(1-\bar{n})  + n_L (1-n_R) \gamma_L \gamma_R\left(1 +  \xi^2\right) }{4g^2 +  \gamma_L \gamma_R(1+ \xi^2)},\\
        &p_R 
        = \frac{4g^2 \bar{n}(1-\bar{n})  + (1-n_L) n_R \gamma_L \gamma_R\left(1 +  \xi^2\right) }{4g^2 +  \gamma_L \gamma_R(1+ \xi^2)},\\
        &\alpha 
        = \frac{2ig(n_L - n_R)\gamma_L \gamma_R(1 + i\xi)}{(\gamma_L + \gamma_R)(4g^2 +  \gamma_L \gamma_R (1+\xi^2))},\\
    \end{split}
\end{equation}
where
\begin{equation}
    \bar{n} = \frac{n_L \gamma_L + n_R \gamma_R}{\gamma_L + \gamma_R}
    \hspace{0.7cm}
    \text{and}
    \hspace{0.5cm}
    \xi = \frac{4\delta}{\gamma_L + \gamma_R}.
\end{equation}

Using the formula for the concurrence, given by Eq.~\eqref{eq:Concurrence}, we obtain the following expression,
\begin{equation}
\label{eq:Concurrence Local e}
    \mathcal{C} = 2 |\alpha| - \frac{2 \gamma_L \gamma_R}{4g^2 + \gamma_L \gamma_R(1 + \xi^2)}  \sqrt{\left[(1-n_L)(1-n_R)(1+\xi^2)  +\frac{4g^2}{\gamma_L \gamma_R}(1-\bar{n})^2 \right]\left[ n_L n_R (1+\xi^2)  + \frac{4g^2}{\gamma_L \gamma_R}\bar{n}^2 \right]},
\end{equation}
\end{widetext}
if this quantity is positive and zero otherwise.

\subsubsection{Average current}
In the steady state, the average current passing through the system from the left reservoir can be obtained from the expression given in Eq.~\eqref{eq: Current SS}.
Equivalently, one may compute the current as an expectation value of the current operator,
\begin{equation}
    \hat{I} = ig(\hat{c}_L^{\dagger} \hat{c}_R - \hat{c}_R^{\dagger} \hat{c}_L).
\end{equation}
In the steady state, we find
\begin{equation}
\label{eq:Current local e}
    \langle I \rangle = \frac{4g^2 (n_L - n_R) \gamma_L \gamma_R (\gamma_L + \gamma_R)}{(\gamma_L + \gamma_R)^2(4g^2 + \gamma_L \gamma_R) + 16 \gamma_L \gamma_R \delta^2}.
\end{equation}

\subsubsection{Heat current}
We define the heat currents in the thermodynamically consistent framework~\cite{Potts_2021, Trushechkin_2021}, which guarantees the thermodynamically consistent heat flow and maintains the accuracy of the local form of the master equation. In this framework the average heat current from the reservoir $\ell$ to the double quantum dot is given by~\cite{Potts_2021}
\begin{equation}
\label{eq:Heat local}
     J_\ell = \text{Tr}\left[(\hat{H}_{TD} - \mu_\ell \hat{N}) \mathcal{L}_\ell \hat{\rho} \right],
\end{equation}
where we have introduced a thermodynamic bookkeeping Hamiltonian
\begin{equation}
    \hat{H}_{TD} = \bar{\epsilon}(\hat{c}_L^{\dagger} \hat{c}_L + \hat{c}_R^{\dagger} \hat{c}_R).
\end{equation}
We find
\begin{equation}
    J_L  = \text{Tr}\left[\left(\bar{\epsilon} - \mu_L \right) \hat{N} \mathcal{L}_L \hat{\rho} \right] = \left(\bar{\epsilon} - \mu_L\right)\langle I \rangle ,
\end{equation}
and
\begin{equation}
   J_R  = \text{Tr}\left[\left(\bar{\epsilon} - \mu_R \right) \hat{N} \mathcal{L}_R \hat{\rho} \right] = - \left(\bar{\epsilon} - \mu_R\right)\langle I \rangle.
\end{equation} 
We note that if $\epsilon_L = \epsilon_R$, which is assumed in the main text of this paper, the steady-state heat currents reduce to the ones obtained from the conventional definition, where in Eq.~\eqref{eq:Heat local} the Hamiltonian $\hat{H}$ is used instead of $\hat{H}_{TD}$.

\subsubsection{Current fluctuations}
In order to investigate the fluctuations of the current, we resort to the method of full counting statistics. All the details about this procedure are explained in Appendix~\ref{App:FCS}.

The counting-field-dependent Liouvillian $\mathcal{L}(\chi)$, introduced in Eq.~\eqref{eq:ME chi}, is obtained by inserting the terms $e^{\pm i \chi}$ into the Lindblad superoperator [Eq.~\eqref{eq:Locale}] of the local master equation,
\begin{widetext}
\begin{equation}
    \label{eq:Local chi}
        \mathcal{L}(\chi)\hat{\rho}(\chi, t) = \mathcal{L}\hat{\rho}(\chi, t) 
    + \gamma_L n_L (e^{i\chi} - 1) \hat{c}_L^{\dagger}\hat{\rho}(\chi, t) \hat{c}_L 
        + \gamma_L(1-n_L) (e^{-i\chi} - 1)\hat{c}_L\hat{\rho}(\chi, t) \hat{c}_L^{\dagger},
\end{equation}
where $\mathcal{L}$ is the Liouvillian in the absence of a counting field given in Eq.~\eqref{eq:LME}.
Writing $\mathcal{L}(\chi)$ in the basis $\{p_0, p_L, p_R, p_D, \text{Re}[\alpha], \text{Im}[\alpha]\}$ corresponding to the elements of the density matrix [cf.~Eq.~\eqref{eq:Density Matrix}] yields
\begin{equation}
\label{eq:Li Local}
\mathbb{L}(\chi) =
\footnotesize
\begin{pmatrix}
        -n_L \gamma_L - n_R \gamma_R & e^{-i\chi}(1-n_L)\gamma_L & (1-n_R)\gamma_R & 0 & 0 & 0 \\
        e^{i \chi} n_L \gamma_L & -(1-n_L)\gamma_L - n_R \gamma_R & 0 & (1-n_R)\gamma_R & 0 & -2g \\
        n_R \gamma_R & 0 & - n_L \gamma_L - (1-n_R) \gamma_R & e^{-i \chi}(1-n_L)\gamma_L & 0 & 2g \\
        0 & n_R \gamma_R & e^{i\chi}n_L \gamma_L & -(1-n_L)\gamma_L - (1-n_R) \gamma_R & 0 & 0 \\
        0 & 0 & 0 & 0 & -\frac{\gamma_L + \gamma_R}{2} & 2\delta \\
        0 & g & -g & 0 & -2\delta & -\frac{\gamma_L + \gamma_R}{2}
    \end{pmatrix}
    \normalsize
    .
\end{equation}
The expression for the current fluctuations is given by
\begin{equation}
\label{eq:Variance local e}
    \langle\!\langle I^2 \rangle\!\rangle = \langle I \rangle \frac{n_L + n_R - 2n_L n_R}{n_L - n_R} 
    - \frac{ 2 \langle I \rangle^2}{\gamma_L + \gamma_R}\left(1+ \frac{(\gamma_L + \gamma_R)^2 + 16\delta^2(\gamma_L - \gamma_R)^2/(\gamma_L + \gamma_R)^2}{4g^2 + \gamma_L \gamma_R + 16 \delta^2 \gamma_L \gamma_R /(\gamma_L + \gamma_R)^2} \right).
\end{equation}
\end{widetext}

\subsection{Global master equation}
\label{App: Global}
In the global master equation electrons jump into the eigenmodes of the Hamiltonian [cf.~\eqref{eq:He}], which can be written in diagonal form,
\begin{equation}
    \label{eq:H Diagonal}
    \hat{H} = \epsilon_+ \hat{c}_+^{\dagger} \hat{c}_+ + \epsilon_- \hat{c}_-^{\dagger} \hat{c}_-,
\end{equation}
where we have introduced the annihilation operators
\begin{equation}
\label{eq:Diagonal Operators}
\begin{split}
        &\hat{c}_+ = \cos{(\theta/2)}\hat{c}_L + \sin{(\theta/2)}\hat{c}_R,\\
        &\hat{c}_- = -\sin{(\theta/2)}\hat{c}_L + \cos{(\theta/2)}\hat{c}_R,
\end{split}
\end{equation}
and the corresponding eigenenergies
\begin{equation}
\label{eq:Eigenenergies}
    \epsilon_{\pm} = \bar{\epsilon} \pm \sqrt{\delta^2 + g^2}.
\end{equation}
Here $\bar{\epsilon} = (\epsilon_L + \epsilon_R)/2$, and the angle $\theta$ ($0 \leq \theta \leq \pi$) is defined by $\cos{(\theta)} = \delta/\sqrt{g^2 + \delta^2}$. The Lindblad superoperators in the global master equation are given by \cite{Potts_2021}
\begin{equation}
\label{eq:Globale}
    \mathcal{L}_\ell = \sum_{s= \pm}  n_{\ell}^s \gamma_{\ell}^s \mathcal{D}[\hat{c}_s^{\dagger}] +  (1 - n_{\ell}^s) \gamma_{\ell}^s \mathcal{D}[\hat{c}_s],
\end{equation}
where we have introduced the rescaled coupling rates
\begin{equation}
\label{eq:Gamma Global}
\begin{split}
    & \gamma_{L}^+ = \gamma_L \cos^2{(\theta/2)}\text{, }\gamma_{L}^- = \gamma_L \sin^2{(\theta/2)},\\
    & \gamma_{R}^+ = \gamma_R \sin^2{(\theta/2)}\text{, }\gamma_{R}^- = \gamma_R \cos^2{(\theta/2)},
\end{split}
\end{equation}
and 
\begin{equation}
\label{eq:Fermi Global}
    n_{\ell}^s := n_{\ell}(\epsilon_s).
\end{equation}
From the global master equation, we find the equations of motion
\begin{equation}
    \frac{d}{dt} \langle \hat{c}_s^{\dagger} \hat{c}_s \rangle = (\gamma_{L}^s n_{L}^s + \gamma_{R}^s n_{R}^s)- (\gamma_{L}^s + \gamma_{R}^s)\langle \hat{c}_s^{\dagger} \hat{c}_s \rangle.
\end{equation}
The corresponding steady-state solution is given by
\begin{equation}
    \langle \hat{c}_s^{\dagger} \hat{c}_s \rangle = \frac{\gamma_{L}^s n_{L}^s + \gamma_{R}^s n_{R}^s}{\gamma_{L}^s + \gamma_{R}^s} =: \bar{n}_s,
\end{equation}
and the remaining expectation values vanish, i.e., $\langle \hat{c}_+^{\dagger} \hat{c}_- \rangle = \langle \hat{c}_-^{\dagger} \hat{c}_+ \rangle = 0$.

The density matrix elements in the local basis can be obtained from Eq.~\eqref{eq:Coefficients}, using the relations
\begin{equation}
    \begin{split}
        &\langle \hat{c}_L^{\dagger} \hat{c}_L \rangle = \sin^2(\theta/2)\langle \hat{c}_-^{\dagger} \hat{c}_- \rangle + \cos^2(\theta/2) \langle \hat{c}_+^{\dagger} \hat{c}_+ \rangle, \\
        &\langle \hat{c}_R^{\dagger} \hat{c}_R \rangle = \cos^2(\theta/2) \langle \hat{c}_-^{\dagger} \hat{c}_- \rangle + \sin^2(\theta/2) \langle \hat{c}_+^{\dagger} \hat{c}_+ \rangle,\\
        &\langle \hat{c}_L^{\dagger} \hat{c}_R \rangle = \sin{(\theta/2)} \cos{(\theta/2)} \left(\langle \hat{c}_+^{\dagger} \hat{c}_+ \rangle - \langle \hat{c}_-^{\dagger} \hat{c}_- \rangle \right),\\
        &\langle \hat{c}_R^{\dagger} \hat{c}_L \rangle = \sin{(\theta/2)} \cos{(\theta/2)} \left(\langle \hat{c}_+^{\dagger} \hat{c}_+ \rangle - \langle \hat{c}_-^{\dagger} \hat{c}_- \rangle \right).
    \end{split}
\end{equation}
The coherence then reads
\begin{equation}
    \label{eq:Coherence Global}
|\alpha| = \frac{g}{2\sqrt{g^2 + \delta^2}}|\bar{n}_+ - \bar{n}_-|,
\end{equation}
whereas the concurrence is given by
\begin{equation}
\label{eq:Concurrence Global}
\begin{split}
\mathcal{C} =  2 |\alpha| - 2 \sqrt{\bar{n}_+ \bar{n}_- (1 - \bar{n}_+) (1-\bar{n}_-)}.
\end{split}
\end{equation}

\subsubsection{Average current}
As for the local master equation, the average may be obtained from Eq.~\eqref{eq: Current SS}.
We note that for the global master equation, we may not use a current operator \cite{hofer_2017lvg}. We find
\begin{equation}
\label{eq:Current global}
    \langle I \rangle = \frac{(n_L^+ - n_R^+)\gamma_L^+ \gamma_R^+}{\gamma_L^+ + \gamma_R^+} + \frac{(n_L^- - n_R^-)\gamma_L^- \gamma_R^-}{\gamma_L^- + \gamma_R^-}.
\end{equation}

\subsubsection{Heat current}

The heat current from reservoir $\ell$ may be written as \cite{Potts_2021}
\begin{equation}
\label{eq:Heat global}
    J_\ell = \text{Tr}\left[ \left( H - \mu_\ell \hat{N} \right) \mathcal{L}_\ell \hat{\rho} \right],
\end{equation}
resulting in
\begin{equation}
    J_\ell = \sum_{s = \pm}\frac{(n_{\ell}^s - n_{\bar{\ell}}^s )\gamma_{\ell}^s \gamma_{\bar{\ell}}^s }{\gamma_{\ell}^s + \gamma_{\bar{\ell}}^s}(\epsilon_s - \mu_\ell),
\end{equation}
where $\bar{\ell} \neq \ell$.

\subsubsection{Current fluctuations}
Similarly to the local master equation, we find a $\chi$-dependent Liouvillian:
\begin{widetext}
\begin{equation}
    \label{eq:Global chi}
        \mathcal{L}(\chi)\hat{\rho}(\chi, t) = \mathcal{L}\hat{\rho}(\chi, t) 
        +\sum_{s = \pm} (e^{i\chi} - 1)\gamma_{L}^s n_{L}^s \hat{c}_s^{\dagger}\hat{\rho}(\chi, t) \hat{c}_s 
        + \sum_{s = \pm}(e^{-i\chi} - 1)\gamma_{L}^s (1 - n_{L}^s)\hat{c}_s\hat{\rho}(\chi, t) \hat{c}_s^{\dagger}.
\end{equation}
In the global master equation, the natural choice of the basis is given by $\{p_0, p_+, p_-, p_D\}$, where $p_\pm = \langle0, 0| \hat{c}_\pm \hat{\rho} \hat{c}_\pm^{\dagger} |0, 0\rangle$. In this basis the Liouvillian reads

\begin{equation}
\label{eq:Li Global}
    \mathbb{L}(\chi) =
    \footnotesize
    \begin{pmatrix}
    -\sum_{s, \ell} n_l^s \gamma_\ell^s & e^{-i\chi} (1-n_L^+)\gamma_L^+ +(1-n_R^+)\gamma_R^+ & e^{-i\chi} (1-n_L^-)\gamma_L^- +(1-n_R^-)\gamma_R^- & 0\\
    e^{i\chi}n_L^+\gamma_L^+ + n_R^+\gamma_R^+ & -\sum_\ell \left[(1-n_\ell^+)\gamma_\ell^+ + n_\ell^- \gamma_\ell^- \right] & 0 &  e^{-i\chi} (1-n_\ell^-)\gamma_L^- +(1-n_R^-)\gamma_R^- \\
    e^{i\chi}n_L^-\gamma_L^- + n_R^-\gamma_R^- & 0 & -\sum_\ell \left[(1-n_\ell^-)\gamma_\ell^- + n_\ell^+ \gamma_\ell^+ \right] &  e^{-i\chi} (1-n_L^+)\gamma_L^+ +(1-n_R^+)\gamma_R^+ \\
    0 & e^{i\chi}n_L^-\gamma_L^- + n_R^-\gamma_R^- & e^{i\chi}n_L^+\gamma_L^+ + n_R^+\gamma_R^+ &  -\sum_{s, \ell} (1-n_\ell^s) \gamma_\ell^s \\
    \end{pmatrix}
    \normalsize
    .
\end{equation}
Using the method of full counting statistics outlined in Appendix~\ref{App:FCS}, we find
\begin{equation}
\label{eq:Variance global}
    \langle\!\langle I^2 \rangle\!\rangle = \sum_{s = \pm} \frac{\gamma_L^s \gamma_R^s}{\gamma_L^s + \gamma_L^s}\left(n_L^s + n_R^s - 2 n_L^s n_R^s-\frac{ 2\gamma_L^s \gamma_R^s \left( n_L^s - n_R^s \right)^2 }{(\gamma_L^s + \gamma_L^s)^2} \right).
\end{equation}
\end{widetext}

\subsection{Nonequilibrium Green's functions}

\label{App:NEGFs}

The key building blocks in the steady-state description with NEGFs are the retarded $\Gb^r(\omega)$ and the lesser $\Gb^<(\omega)$
Green's functions \cite{Bruus, Meir_1992}:
\begin{align}
  &\Gb^r(\omega) = 1 /\left(\omega {\bm 1} - \mathbf{H} - \Sigmab^r(\omega)\right), \label{eq:retG}\\
&\Gb^<(\omega) = \Gb^r(\omega) \Sigmab^< (\omega) \Gb^a(\omega). \label{eq:lesserG}
\end{align}
We use boldface symbols to represent  $2 \times 2$ matrices in the site indices (L, R) of the QD system. ${\bm 1} $ is the identity matrix, the Hamiltonian of the system [cf.~Eq.~\eqref{eq:He}] is represented by
\begin{equation}
    \mathbf{H} =
    \begin{pmatrix}
        \epsilon_L & g\\
        g & \epsilon_R\\
    \end{pmatrix},
\end{equation}
and $\Gb^a = (\Gb^r)^\dagger$. In the absence of Coulomb interaction the retarded or lesser self-energy $\Sigmab^{r/<} (\omega)$ describes the coupling between the double quantum dot and the reservoirs. In the wideband limit we have~\cite{Jauho_1994, Meir_1992, Myohanen_2008}
\begin{equation}
    \Sigmab^{<} (\omega) = i
    \begin{pmatrix}
        n_L(\omega) \gamma_L & 0\\
        0 & n_R(\omega) \gamma_R\\
    \end{pmatrix},
\end{equation}
and
\begin{equation}
    \Sigmab^{r} (\omega) = -\frac{i}{2}
    \begin{pmatrix}
        \gamma_L & 0\\
        0 & \gamma_R\\
    \end{pmatrix}.
\end{equation}



From Eqs.~\eqref{eq:retG} and \eqref{eq:lesserG} we find the following four elements of $\Gb^<(\omega)$, given by
\begin{widetext}
\begin{equation}
\label{eq:G lesser}
    \begin{split}
    & G_{LL}^<(\omega) = i\frac{|\omega - \epsilon_R + \frac{i\gamma_R}{2}^2\gamma_L n_L(\omega) + g^2 \gamma_R n_R(\omega) }{|(\omega - \epsilon_R + \frac{i\gamma_R}{2})(\omega - \epsilon_L + \frac{i\gamma_L}{2}) - g^2|^2},\\
    & G_{RR}^<(\omega) = i\frac{|\omega - \epsilon_L + \frac{i\gamma_L}{2}^2\gamma_R n_R(\omega) + g^2 \gamma_L n_L(\omega) }{|(\omega - \epsilon_R + \frac{i\gamma_R}{2})(\omega - \epsilon_L + \frac{i\gamma_L}{2}) - g^2|^2},\\
    & G_{LR}^<(\omega) = ig\frac{(\omega - \epsilon_R + \frac{i\gamma_R}{2})\gamma_L n_L(\omega) + (\omega - \epsilon_L - \frac{i\gamma_L}{2}) \gamma_R n_R(\omega) }{|(\omega - \epsilon_L + \frac{i\gamma_L}{2})(\omega - \epsilon_R + \frac{i\gamma_R}{2})) - g^2|^2},\\
    & G_{RL}^<(\omega) = ig\frac{(\omega - \epsilon_R - \frac{i\gamma_R}{2})\gamma_L n_L(\omega) + (\omega - \epsilon_L + \frac{i\gamma_L}{2}) \gamma_R n_R(\omega) }{|(\omega - \epsilon_L + \frac{i\gamma_L}{2})(\omega - \epsilon_R + \frac{i\gamma_R}{2})) - g^2|^2}.\\
\end{split}
\end{equation}
\end{widetext}
The expectation values $\langle \hat{c}_\ell^{\dagger} \hat{c}_{\ell^{'}} \rangle$ ($\ell, \ell^{'} = L, R$) are obtained by taking the inverse Fourier transform of the corresponding lesser Green's functions \cite{Bruus, Meir_1992}
\begin{equation}
    \langle \hat{c}_\ell^{\dagger} \hat{c}_{\ell^{'}} \rangle = -i\mathcal{F}\left[G_{\ell^{'} \ell}^<(\omega)\right],
\end{equation}
where we use the convention
\begin{equation}
    \mathcal{F}\left[G(\omega)\right] = \lim_{\tau \to 0} \frac{1}{2 \pi} \int_{-\infty}^{\infty}  d \omega e^{-i \omega \tau} G(\omega).
\end{equation}

Cumulants of the electron current through the system can be obtained from the transmission function $\mathcal{T}(\omega)$. The expression for the average current reads \cite{Levitov_1993, Levitov_1996, Levitov_2004}
\begin{equation}
    \langle I \rangle = \int_{-\infty}^{\infty} \frac{d\omega}{2 \pi} \mathcal{T}(\omega) \left[n_L(\omega) - n_R(\omega) \right],
\end{equation}
and the fluctuations are given by
\begin{equation}
\begin{split}
    \langle\!\langle I^2& \rangle\!\rangle = \int_{-\infty}^{\infty} \frac{d\omega}{2 \pi} \mathcal{T}(\omega) \{n_L(\omega) + n_R(\omega)\\& -2 n_L(\omega) n_R(\omega) 
    - \mathcal{T}(\omega)[n_L(\omega) - n_R(\omega)]^2 \}.
\end{split}
\end{equation}
The transmission function itself can be expressed with the Green's functions~\cite{Meir_1992}
\begin{equation}
    \mathcal{T}(\omega) = \text{Tr} \left[\Gb^a (\omega) \Gammab_L \Gb^r (\omega) \Gammab_R  \right],
\end{equation}
where
\begin{equation}
    \Gammab_L = \text{Diag}\left(\gamma_L, 0 \right) \hspace{0.5cm} \text{and} \hspace{0.5cm} \Gammab_R = \text{Diag}\left(0, \gamma_R \right).
\end{equation}
For the double quantum dot system it is given by \cite{Sumetskii_1993, Agarwalla_2018}
\begin{equation}
    \label{eq:Transmission function e}
    \mathcal{T}(\omega) = \frac{\gamma_L \gamma_R g^2}{|\left(\omega - \epsilon_L + i\frac{\gamma_L}{2} \right) \left(\omega - \epsilon_R + i\frac{\gamma_R}{2} \right)-g^2|^2}.
\end{equation}
In the same framework, the heat current from reservoir $\ell$ reads
\begin{equation}
     J_\ell = \int_{-\infty}^{\infty} \frac{d\omega}{2 \pi} (\omega - \mu_\ell)\mathcal{T}(\omega) \left[n_\ell(\omega) - n_{\bar{\ell}}(\omega) \right],
\end{equation}
where $\bar{\ell} \neq \ell$.

\subsection{Stochastic model of the system}
\label{App:Stochastic model}
To motivate the classical model, given by the rate equation in Eq.~\eqref{eq:SM}, we consider the local master equation
\begin{equation}
\label{eq:SM1}
    \frac{d \hat{\rho}(t)}{dt} = \mathcal{L}\hat{\rho}(t),
\end{equation}
with $\mathcal{L}$ given in Eqs.~\eqref{eq:LME} and \eqref{eq:Local}.

We now introduce Nakajima-Zwanzig superoperators $\mathcal{P}$ and $\mathcal{Q} = \mathcal{I} - \mathcal{P}$, such that $\mathcal{P}$ projects onto the population subspace of $\hat{\rho}$. 
Here $\mathcal{I}$ is the identity superoperator. We note that $\mathcal{P}$ and $\mathcal{Q}$ are projectors, i.e., $\mathcal{P}^2 = \mathcal{P}$, $\mathcal{Q}^2 = \mathcal{Q}$, and $\mathcal{Q}\mathcal{P} = 0$. We also decompose the Liouvillian of the system into $\mathcal{L} = \mathcal{L}_0 + \mathcal{V}$, where $\mathcal{V}$ is the contribution due to the inter-dot tunneling, i.e., $\mathcal{V}\hat{\rho} = -ig[\hat{c}_L^{\dagger} \hat{c}_R + \hat{c}_R^{\dagger} \hat{c}_L, \hat{\rho}]$. For small $g$ we can treat $\mathcal{V}$ as a perturbation. In the following derivation, we will use the relations
\begin{equation}
\label{eq:SM2}
    \mathcal{P} \mathcal{V} \mathcal{P} = [\mathcal{L}_0, \mathcal{P}] = [\mathcal{L}_0, \mathcal{Q}] = 0.
\end{equation}

By acting with $\mathcal{P}$ and $\mathcal{Q}$ on Eq.~\eqref{eq:SM1} we obtain the equations
\begin{equation}
\label{eq:SM3}
     \frac{d \mathcal{P}\hat{\rho}(t)}{dt} = \mathcal{P}\mathcal{L} \mathcal{P}\hat{\rho}(t) + \mathcal{P}\mathcal{L} \mathcal{Q}\hat{\rho}(t)
\end{equation}
and
\begin{equation}
\label{eq:SM4}
     \frac{d \mathcal{Q}\hat{\rho}(t)}{dt} = \mathcal{Q}\mathcal{L} \mathcal{P}\hat{\rho}(t) + \mathcal{Q}\mathcal{L} \mathcal{Q}\hat{\rho}(t).
\end{equation}
The formal solution to the second equation reads
\begin{equation}
    \label{eq:SM5}
    \mathcal{Q}\hat{\rho}(t) = \mathcal{G}(t, 0) \mathcal{Q} \hat{\rho}(0) + \int_0^t ds \mathcal{G}(t, s) \mathcal{Q} \mathcal{L} \mathcal{P} \hat{\rho}(s),
\end{equation}
where we have introduced the propagator
\begin{equation}
\label{eq:SM6}
    \mathcal{G}(t, s) = e^{\int_s^t d \tau \mathcal{Q} \mathcal{L} }.
\end{equation}
After inserting Eq.~\eqref{eq:SM5} into Eq.~\eqref{eq:SM3} and assuming that $\mathcal{Q} \hat{\rho}(0) = 0$, we find
\begin{equation}
\label{eq:SM7}
    \frac{d \mathcal{P}\hat{\rho}(t)}{dt} = \mathcal{P}\mathcal{L} \mathcal{P}\hat{\rho}(t) + \mathcal{P}\mathcal{L} \int_0^t ds \mathcal{G}(t, s) \mathcal{Q} \mathcal{L} \mathcal{P} \hat{\rho}(s).
\end{equation}
By inserting $\mathcal{L} = \mathcal{L}_0 + \mathcal{V}$, applying Eq.~\eqref{eq:SM2}, and using the properties of Nakajima-Zwanzig projectors, we obtain the formal solution
\begin{equation}
\label{eq:SM8}
\begin{split}
    &\frac{d \mathcal{P}\hat{\rho}(t)}{dt} = \mathcal{L}_0 \mathcal{P} \hat{\rho}(t) \\
    &+(\mathcal{P} \mathcal{V} \mathcal{Q}) \int_0^t ds e^{(\mathcal{Q} \mathcal{L}_0 \mathcal{Q} + \mathcal{Q} \mathcal{V} \mathcal{Q})s}(\mathcal{Q} \mathcal{V} \mathcal{P})\mathcal{P} \hat{\rho}(t-s),
\end{split}
\end{equation}
where we have also made a substitution $s \to t-s$. The eigenvalues of $\mathcal{V}$ and $\mathcal{L}_0$ are proportional to $g$ and $\gamma_{\ell}$, respectively. For $g \ll \gamma_{\ell}$, we can treat $\mathcal{V}$ as a perturbation and neglect its contribution to the exponential $e^{(\mathcal{Q} \mathcal{L}_0 \mathcal{Q} + \mathcal{Q} \mathcal{V} \mathcal{Q})s}$. This exponential thus decays on the timescale $1/\gamma_{\ell}$. We now assume that the diagonal elements of the density matrix remain approximately constant during this timescale, an assumption which is not justified (see below for more information). This allows us to substitute $\mathcal{P}\hat{\rho}(t)$ for $\mathcal{P}\hat{\rho}(t-s)$ in Eq.~\eqref{eq:SM8} and to extend the integral to infinity. Equation \eqref{eq:SM8} then reduces to a Markovian master equation for $\mathcal{P}\hat{\rho}$,
\begin{equation}
    \label{eq:SM9}
     \frac{d\mathcal{P}\hat{\rho}(t)}{dt} = (\mathcal{L}_0 - \mathcal{P} \mathcal{V} \mathcal{Q} \mathcal{L}_0^{-1} \mathcal{Q} \mathcal{V} \mathcal{P})\mathcal{P}\hat{\rho}(t),
\end{equation}
where
\begin{equation}
    \label{eq:SM10}
    \mathcal{L}_0^{-1} = - \int_0^{\infty} ds e^{\mathcal{L}_0 s} \mathcal{Q}
\end{equation}
is the Drazin inverse of $\mathcal{L}_0$.

\begin{widetext}
Equation \eqref{eq:SM9} can be cast into the form given in Eq.~\eqref{eq:SM} with the stochastic matrix
\begin{equation}
\label{eq:W}
W =
\footnotesize
\begin{pmatrix}
        -n_L \gamma_L - n_R \gamma_R & (1-n_L)\gamma_L & (1-n_R)\gamma_R & 0 \\
        n_L \gamma_L & -(1-n_L)\gamma_L - n_R \gamma_R-\frac{4g^2}{\gamma_L + \gamma_R} & \frac{4g^2}{\gamma_L + \gamma_R} & (1-n_R)\gamma_R \\
        n_R \gamma_R & \frac{4g^2}{\gamma_L + \gamma_R}& - n_L \gamma_L - (1-n_R)\gamma_R -\frac{4g^2}{\gamma_L + \gamma_R} \gamma_R & (1-n_L)\gamma_L  \\
        0 & n_R \gamma_R & n_L \gamma_L & -(1-n_L)\gamma_L - (1-n_R) \gamma_R  
    \end{pmatrix}
    \normalsize
    .
\end{equation}
\end{widetext}
All entries in $W$ are identical to the ones in the local master equation [cf.~\eqref{eq:Li Local}] except the element $W_{LR} = W_{RL}$,
which comes from the term $\mathcal{P} \mathcal{V} \mathcal{Q} \mathcal{L}_0^{-1} \mathcal{Q} \mathcal{V} \mathcal{P}$ (here we assumed $\delta=0$ for simplicity).

We stress that the approximation giving rise to Eq.~\eqref{eq:SM9} is not justified. It assumes that the diagonal elements remain constant over the timescale $1/\gamma_{ \ell}$. This is, however, exactly the timescale over which the diagonal elements do change. For our purposes, this is not problematic since our aim is to derive an analogous classical description for the system. We do not claim that this classical analog is exact in any limit.

Interestingly, we find the steady-state populations in the classical model to be the same as in the local master equation given in Eq.~\eqref{eq:Solution local} for all choices of parameters. The same holds for the average current, which agrees with expression \eqref{eq:Current local} obtained with the local master equation. The fact that the derived rate equation reproduces the average behavior of the local master equation further justifies its use as an analogous classical model.

To obtain the current fluctuations, counting fields may be introduced in analogy to Eq.~\eqref{eq:Li Local}. Using full counting statistics, we then obtain
\begin{equation}
    \label{Variance sm}
    \begin{split}
       & \langle\!\langle I^2 \rangle\!\rangle = \frac{4g^2 \gamma_L \gamma_R (n_L(1-n_R) + n_R(1-n_L))}{(\gamma_L + \gamma_R)(4g^2 + \gamma_L \gamma_R)} \\
        &- \frac{32[g^2 \gamma_L \gamma_R (n_L - n_R)]^2[4g^2 - \gamma_L \gamma_R + (\gamma_L + \gamma_R)^2]}{[(\gamma_L + \gamma_R)(4g^2 + \gamma_L \gamma_R)]^3}.
    \end{split}
\end{equation}

\section{Full counting statistics}
\label{App:FCS}

In order to obtain the expressions for the current fluctuations, we employ the technique of full counting statistics. In this framework the density matrix of the system is resolved into $\hat{\rho}(n, t)$, the (unnormalized) density matrices given that a net number of $n$ electrons tunneled from the left reservoir into the DQD during the time interval $[0, t]$ \cite{Esposito_2009, Plenio_1998, Schaller_2013}.
The probability that $n$ electrons have tunneled can then be obtained by 
\begin{equation}
    P[n] = \text{Tr}[\hat{\rho}(n, t)].
\end{equation}
The original density matrix can be recovered by taking the sum
\begin{equation}
\label{eq: Rho n}
    \hat{\rho}(t) = \sum_n \hat{\rho}(n, t).
\end{equation}

The cumulants of the net number of transported electrons can be obtained from the cumulant generating function $S(\chi, t)$, given by
\begin{equation}
    S(\chi, t) =  \log{ \sum_n P[n]e^{i\chi n}} = \log{  \text{Tr}\left[ \hat{\rho}(\chi, t)\right]},
\end{equation}
where 
\begin{equation}
\label{eq: Rho chi}
    \hat{\rho}(\chi, t) = \sum_n \hat{\rho}(n, t) e^{i  \chi n},
\end{equation}
and $\chi$ is known as the \textit{counting field}.

The time evolution of $\hat{\rho}(\chi, t)$ can be found by inserting the $\chi$-dependent term, which tracks electron jumps, into the original Lindblad master equation \cite{Schaller_2013}. This results in a counting-field-dependent master equation,
\begin{equation}
    \label{eq:ME chi}
    \frac{d \hat{\rho}(\chi, t)}{dt} = \mathcal{L}(\chi)\hat{\rho}(\chi, t).
\end{equation}
The specific form of $\mathcal{L}(\chi)$ depends on the choice of the master equation, and the original Liouvillian can be recovered with $\mathcal{L} = \mathcal{L}(0)$.
The formal solution to Eq.~\eqref{eq:ME chi} is given by
\begin{equation}
    \hat{\rho}(\chi, t) = e^{\mathcal{L}(\chi) t }\hat{\rho}(0, 0).
\end{equation}

The cumulants of the electron current in the steady state can be obtained from the scaled cumulant generating function of $n$ in the long-time limit,
    \begin{equation}
    \label{eq:cumulants}
    \langle \! \langle I^k \rangle\! \rangle = \frac{d^k}{d(i\chi)^k}\left(\lim_{t \rightarrow \infty} \frac{S(\chi, t)}{t}\right)|_{\chi = 0} = \frac{d^k}{d(i\chi)^k} \lambda(\chi)|_{\chi = 0},
\end{equation}
where $\lambda(\chi)$ is the eigenvalue of $\mathcal{L}(\chi)$ with the largest real part. The average of the current, $\langle I \rangle = \langle\!\langle I \rangle\!\rangle$, and its fluctuations, $\langle\!\langle I^2 \rangle\!\rangle$, can be evaluated by taking $k = 1, 2$, respectively.

Since a direct of computation of $\lambda(\chi)$ is not straightforward for our system, we make use of the scheme described in~\cite{Bruderer_2014}. We express the Liouvillian $\mathcal{L}(\chi)$ as a matrix $\mathbb{L}(\chi)$ in a basis that can be freely chosen and may be different depending on whether we use the local or global master equation. Let $P_{\chi}(x)$ be the characteristic polynomial
\begin{equation}
    P_{\chi}(x) = \det{\left[\mathbb{L}(\chi) - x \mathbb{I} \right]} ,
\end{equation}
where $\mathbb{I}$ is the identity matrix. Upon expanding in $x$ we obtain
    \begin{equation}
    \label{eq:P chi expansion}
    P_{\chi}(x)  = \sum_{j=0}^M  a_j x^j = \sum_{j= 0}^M \sum_{k= 0}^{\infty}  \frac{a_j^{(k)} (i\chi)^k}{k!}x^j,
\end{equation}
where
\begin{equation}
    a_j = \frac{\partial^j}{\partial x^j} \frac{P_{\chi}(x)}{j!}|_{x=0} \hspace{0.4cm} \text{and} \hspace{0.4cm} a_j^{(k)} = \frac{d^k}{d(i\chi)^k} a_j|_{\chi = 0}.
\end{equation}
Here $M$ is the dimension of the matrix $\mathbb{L}(\chi)$. By definition, the eigenvalue $\lambda(\chi)$ of $\mathbb{L}(\chi)$ (with the largest real part) fulfills
\begin{equation}
    P_{\chi}(\lambda(\chi)) = 0.
\end{equation}
It follows that
\begin{equation}
    0 = \frac{d}{d(i\chi)}P_{\chi}(\lambda(\chi))|_{\chi=0} = a_0^{(1)} + a_1^{(0)} \langle I\rangle.
\end{equation}
In order to obtain the right-hand side, we first expanded $P_{\chi}(\lambda(\chi))$ using Eq.~\eqref{eq:P chi expansion}, used $\lambda(0) = 0$, as well as Eq.~\eqref{eq:cumulants}.
We thus find
\begin{equation}
    \langle I \rangle  = -\frac{a_0^{(1)}}{a_1^{(0)}}.
\end{equation}
The second cumulant of the current may be obtained in a similar fashion from 
\begin{equation}
    0 = \frac{d^2}{d(i\chi)^2}P_{\chi}(\lambda(\chi))|_{\chi=0} ,
\end{equation}
resulting in
\begin{equation}
    \langle\!\langle I^2 \rangle\!\rangle =  -\frac{a_0^{(2)} + 2a_1^{(1)} \langle I \rangle + 2 a_2^{(0)} \langle I \rangle^2}{a_1^{(0)}}.
\end{equation}

\section{Interacting electrons}
\label{App: Interacting}
When investigating the system interacting electrons we restrict ourselves to the case with equal onsite energies of quantum dots, where the Hamiltonian of the system is given by Eq.~\eqref{eq:H}. In the presence of Coulomb repulsion, the coefficients of the density matrix in the basis already introduced in Eq.~\eqref{eq:Coefficients} are given by
\begin{equation}
\label{eq:Coefficients U}
    \begin{split}
        &p_0 = 1- \langle \hat{c}_L^{\dagger} \hat{c}_L \rangle - \langle \hat{c}_R^{\dagger} \hat{c}_R \rangle + \langle \hat{c}_L^{\dagger} \hat{c}_L \hat{c}_R^{\dagger} \hat{c}_R \rangle, \\
        &p_L = \langle \hat{c}_L^{\dagger} \hat{c}_L \rangle - \langle \hat{c}_L^{\dagger} \hat{c}_L \hat{c}_R^{\dagger} \hat{c}_R \rangle, \\
        &p_R = \langle \hat{c}_R^{\dagger} \hat{c}_R \rangle - \langle \hat{c}_L^{\dagger} \hat{c}_L \hat{c}_R^{\dagger} \hat{c}_R \rangle, \\
        &p_D=  \langle \hat{c}_L^{\dagger} \hat{c}_L \hat{c}_R^{\dagger} \hat{c}_R \rangle, \\
        &\alpha = \langle \hat{c}_R^{\dagger} \hat{c}_L \rangle.\\
    \end{split}
\end{equation}

\begin{widetext}
\subsection{Semilocal master equation}
\label{App: Semi local}

In the semilocal master equation, the Lindblad superoperator for bath $\ell$ is given by \cite{Potts_2021}
    \begin{equation}
    \label{eq:Semi local}
    \mathcal{L}_\ell = \gamma_\ell \left[n_\ell \mathcal{D}[(1-\hat{c}_{\bar{\ell}}^{\dagger}\hat{c}_{\bar{\ell}}) \hat{c}_\ell^{\dagger}] + n_\ell^U \mathcal{D}[\hat{c}_{\bar{\ell}}^{\dagger}\hat{c}_{\bar{\ell}} \hat{c}_\ell^{\dagger}] + (1-n_\ell) \mathcal{D}[(1-\hat{c}_{\bar{\ell}}^{\dagger}\hat{c}_{\bar{\ell}}) \hat{c}_\ell] + (1-n_\ell^U) \mathcal{D}[\hat{c}_{\bar{\ell}}^{\dagger}\hat{c}_{\bar{\ell}} \hat{c}_\ell] \right],
\end{equation}
where
\begin{equation}
    n_\ell := n_\ell(\epsilon),
\end{equation}
and
\begin{equation}
    n_\ell^U := n_\ell(\epsilon + U).
\end{equation}

The steady state of the system is given by the eigenvector corresponding to the zero eigenvalue of the Liouvillian. TheLiouvillian can be computed from Eq.~\eqref{eq:Li U} evaluated at zero counting field $\chi = 0$. We find
\begin{equation}
\label{eq: coefficients Semi-local}
    \begin{split}
        &p_0 =  \frac{4g^2 (1-\bar{n}) (1-\bar{n}_U) +  \gamma_L \gamma_R (1  -\bar{n} + \bar{n}_U) \left((1-n_L)(1-n_R) - n_L^U n_R^U + \sum_l \frac{\gamma_l n_l^U(n_{\bar{l}} - n_{l} \Delta_{\bar{l}})}{\gamma_L + \gamma_R}\right)}{4g^2 (1 + \bar{n} - \bar{n}_U) + \gamma_L \gamma_R (1-\Delta_L \Delta_R)(1  -\bar{n} + \bar{n}_U)}\\
        &p_L =  \frac{4g^2 \bar{n} (1-\bar{n}_U)  +  \gamma_L \gamma_R (1  -\bar{n} + \bar{n}_U) \left(n_L(1-n_R) + \sum_l \frac{\gamma_l n_l n_l^U  \Delta_{\bar{l}}}{\gamma_L + \gamma_R} + \frac{\gamma_R (n_L^U n_R - n_L n_R^U)}{\gamma_L + \gamma_R}\right)}{4g^2 (1 + \bar{n} - \bar{n}_U)  +  \gamma_L \gamma_R (1-\Delta_L \Delta_R)(1  -\bar{n} + \bar{n}_U)} \\
        &p_R =  \frac{4g^2 \bar{n} (1-\bar{n}_U)  +  \gamma_L \gamma_R (1  -\bar{n} + \bar{n}_U) \left((1-n_L)n_R + \sum_l \frac{\gamma_l n_l n_l^U  \Delta_{\bar{l}}}{\gamma_L + \gamma_R} + \frac{\gamma_L (n_R^U n_L - n_R n_L^U)}{\gamma_L + \gamma_R}\right)}{4g^2 (1 + \bar{n} - \bar{n}_U)  +  \gamma_L \gamma_R (1-\Delta_L \Delta_R)(1  -\bar{n} + \bar{n}_U)} \\
        &p_D = \frac{4g^2 \bar{n} \bar{n}_U  +  \gamma_L \gamma_R (1  -\bar{n} + \bar{n}_U)  \sum_l \frac{\gamma_l n_l^U(n_{\bar{l}} - n_{l} \Delta_{\bar{l}})}{\gamma_L + \gamma_R}}{4g^2 (1 + \bar{n} - \bar{n}_U) + \gamma_L \gamma_R (1-\Delta_L \Delta_R)(1  -\bar{n} + \bar{n}_U)} \\
        &\alpha = \frac{2ig \gamma_L \gamma_R \left[ n_L(1-n_R^U) - n_R(1-n_L^U)\right]}{(\gamma_L + \gamma_R)\left[ 4g^2(1 + \bar{n} - \bar{n}_U ) + (1-\Delta_L \Delta_R)\gamma_L \gamma_R(1 - \bar{n} + \bar{n}_U)\right]}, \\
    \end{split}
\end{equation}
where
\begin{equation}
    \bar{n}_U = \frac{\gamma_L n_L^U + \gamma_R n_R^U}{\gamma_L + \gamma_R},
\end{equation}
and $\Delta_l = n_l - n_l^U$.
Expressions for coherence $|\alpha|$ as well as the concurrence $\mathcal{C}$, which is defined in Eq.~\eqref{eq:Concurrence}, can be obtained from the coefficients shown above.

\subsubsection{Average electron current}
For the average electron current through the system, given by Eq.~\eqref{eq: Current SS}, we find
\begin{equation}
    \label{Current semi local}
    \langle I \rangle = \frac{4g^2 \gamma_L \gamma_R \left[ n_L(1-n_R^U) - n_R(1-n_L^U)\right]}{(\gamma_L + \gamma_R)\left[ 4g^2(1 + \bar{n} - \bar{n}_U ) + (1-\Delta_L \Delta_R)\gamma_L \gamma_R(1 - \bar{n} + \bar{n}_U)\right]}.
\end{equation}

\subsubsection{Heat currents}
The heat current from bath $\ell$ is defined in complete analogy to the local master equation, i.e., Eq.~\eqref{eq:Heat local} \cite{Potts_2021}. In the presence of Coulomb interactions, the bookkeeping Hamiltonian reads
\begin{equation}
    \hat{H}_{TD} =  \epsilon ( \hat{c}_L^{\dagger} \hat{c}_L + \hat{c}_R^{\dagger} \hat{c}_R) + U \hat{c}_L^{\dagger} \hat{c}_L \hat{c}_R^{\dagger} \hat{c}_R,
\end{equation}
resulting in
\begin{equation}
\begin{split}
     \langle Q_L \rangle &= (\epsilon - \mu_L)\langle I \rangle  \\
    & +\frac{U \gamma_L \gamma_R}{\gamma_L + \gamma_R} \frac{4g^2 (n_L^U - n_R^U) \bar{n} + \gamma_L \gamma_R(1- \bar{n} + \bar{n}_U) (n_L^U n_R - n_R^U n_L - n_L n_L^U \Delta_R + n_R n_R^U \Delta_L)}{ 4g^2(1 + \bar{n} - \bar{n}_U ) + (1-\Delta_L \Delta_R)\gamma_L \gamma_R(1 - \bar{n} + \bar{n}_U)},
\end{split}
\end{equation}
and
\begin{equation}
\begin{split}
     \langle Q_R \rangle &= -(\epsilon - \mu_R)\langle I \rangle \\
    & -\frac{U \gamma_L \gamma_R}{\gamma_L + \gamma_R} \frac{4g^2 (n_L^U - n_R^U) \bar{n} + \gamma_L \gamma_R(1- \bar{n} + \bar{n}_U) (n_L^U n_R - n_R^U n_L - n_L n_L^U \Delta_R + n_R n_R^U \Delta_L)}{ 4g^2(1 + \bar{n} - \bar{n}_U ) + (1-\Delta_L \Delta_R)\gamma_L \gamma_R(1 - \bar{n} + \bar{n}_U)},
\end{split}
\end{equation}
respectively.

\subsubsection{Current fluctuations}
The current fluctuations are obtained in complete analogy to the method used in the noninteracting case, i.e., full counting statistics (cf. Appendix~\ref{App:FCS}). The $\chi$-dependent Liouvillian reads
\begin{equation}
    \label{eq:Semi local chi}
    \begin{split}
        \frac{d \hat{\rho}}{dt} & = \mathcal{L}\hat{\rho}  + \gamma_L  (e^{i\chi} - 1)  \left[n_L (1-c_{R}^{\dagger}c_{R}) \hat{c}_L^{\dagger} \hat{\rho} \hat{c}_L (1-c_{R}^{\dagger}c_{R})  + n_L^U (\hat{c}_R^{\dagger}\hat{c}_R) \hat{c}_L^{\dagger} \hat{\rho} \hat{c}_L (\hat{c}_R^{\dagger}\hat{c}_R) \right]\\
        & + \gamma_L  (e^{-i\chi} - 1)  \left[n_L (1-c_{R}^{\dagger}c_{R}) \hat{c}_L \hat{\rho} \hat{c}_L^{\dagger} (1-c_{R}^{\dagger}c_{R})  + n_L^U (\hat{c}_R^{\dagger}\hat{c}_R) \hat{c}_L \hat{\rho} \hat{c}_L^{\dagger} (\hat{c}_R^{\dagger}\hat{c}_R) \right]
    \end{split}
\end{equation}
In the basis $\{p_0, p_L, p_R, p_D, \text{Re}[\alpha], \text{Im}[\alpha]\}$ it is given by 
\begin{equation}
\label{eq:Li U}
    \mathbb{L}(\chi) = 
    \begin{pmatrix}
        -n_L \gamma_L - n_R \gamma_R & e^{-i\chi}(1-n_L)n_L & (1-n_R)\gamma_R & 0 & 0 & 0 \\
        e^{i \chi} n_L \gamma_L & -(1-n_L)\gamma_L - n_R^U \gamma_R & 0 & (1-n_R^U)\gamma_R & 0 & -2g \\
        n_R \gamma_R & 0 & - n_L^U \gamma_L - (1-n_R) \gamma_R & e^{-i \chi}(1-n_L^U)\gamma_L & 0 & 2g \\
        0 & n_R \gamma_R & e^{i\chi}n_L \gamma_L & -(1-n_L)\gamma_L - (1-n_R) \gamma_R & 0 & 0 \\
        0 & 0 & 0 & 0 & \gamma_U & 2\delta \\
        0 & g & -g & 0 & -2\delta & \gamma_U
    \end{pmatrix},
\end{equation}
where
$\gamma_U = -\sum_l \frac{1-n_l+n_l^U}{2}\gamma_l$.
We thereby find the current fluctuations
\begin{equation}
\label{eq:Variance semi local}
\begin{split}
    \langle \langle I^2 \rangle \rangle & =
    \frac{n_L(1-n_R) + n_R(1-n_L) + 2n_L n_R \bar{n}_U - 2 n_L^U n_R^U \bar{n} + \frac{\gamma_L(n_L n_R^U - n_R n_L^U) + \gamma_R (n_R n_L^U -n_L n_R^U)}{\gamma_L + \gamma_R}}{ (n_L(1-n_R^U) -n_R(1-n_L^U))} \langle I \rangle \\
    & + \frac{2 \langle I \rangle^2}{\gamma_L + \gamma_R}
    \left( \frac{n_L^U(1-n_R) - n_R^U (1-n_L) }{ n_L(1-n_R^U) -n_R(1-n_L^U)}  - \frac{2(1 - \Delta_L \Delta_R)\gamma_L \gamma_R + 8g^2 + (\gamma_L + \gamma_R)^2 (1-\bar{n} + \bar{n}_U)}{ 4g^2(1 + \bar{n} - \bar{n}_U ) + (1-\Delta_L \Delta_R)\gamma_L \gamma_R(1 - \bar{n} + \bar{n}_U) } \right).
\end{split}
\end{equation}

\end{widetext}

\subsection{Nonequilibrium Green's functions}
\label{App:NEGF2B}

To determine the densities of the system via NEGFs, we can still proceed as in Appendix~\ref{App:NEGFs}. However, we have to account, within NEGF, for the Coulomb interaction via a many-body contribution to the self-energy \cite{Kadanoff1962,Baym1962,Keldysh1965}. 
As before, the key building blocks in the steady-state description with NEGF in the presence 
of interactions are the retarded $\Gb^r(\omega)$ and lesser $\Gb^<(\omega)$
Green's functions, defined in Eqs.~\eqref{eq:retG} and~\eqref{eq:lesserG} and for convenience repeated here:
\begin{align}
  &\Gb^r(\omega) = 1 /\left(~\omega {\bm 1} -{\bm H} - \Sigmab^r(\omega) \right),\\
&\Gb^<(\omega) = \Gb^r(\omega) \Sigmab^< (\omega) \Gb^a(\omega).
\end{align}
%
The self-energy $\Sigmab$ now has two parts, coming from the embedding (emb) 
and many-body (MB) contributions:
$\Sigmab^{r/<}=\Sigmab^{r/<} _{\text{emb}}+\Sigmab^{r/<} _{\MB}$,
where the embedding term has been discussed before in Appendix.~\ref{App:NEGFs}.
The correlation effects due to interactions are introduced via
the many-body self-energy $\Sigmab_{\MB}^{r/<} = \Sigmab_{\MB}^{r/<} [\Gb^r,\Gb^<]$.
For $\Sigmab_{\MB}$ we use
the second Born approximation \cite{Myohanen_2008,Friesen2009,Karlsson2014b}, keeping all Feynman diagrams up to second order. 
These equations are then solved self-consistently, using 
the Pulay scheme \cite{Pulay1980} to improve convergence. Self-consistency ensures that
general conservation laws \cite{Baym1962,Kadanoff1962,Keldysh1965} are obeyed (e.g., if $I_L$ is the current 
from lead $L$, then $I_R = -I_L $). The embedding self-energy is
independent of the self-consistency cycle and is calculated separately only once.
For the particle density in the device region and the average current $\langle I \rangle$ coming from the left reservoir, the expressions are \cite{Meir_1992}, respectively,
\begin{align}
 & \langle \hat{N}_\ell \rangle = \langle \hat{c}_\ell^{\dagger} \hat{c}_\ell \rangle =  \int _{-\infty} ^\infty \frac{d\omega}{2\pi i} G_{\ell \ell}^<(\omega), \label{eq:density}\\
 &\langle I \rangle =\! \int _{-\infty} ^\infty \frac{d\omega}{2\pi i} 
 \Tr \left [\Gammab_L  \left ( \Gb^< (\omega) - 2\pi i n_L(\omega) {\bm A} (\omega)   \right )     \right ]. \label{eq:meir}
\end{align}
Here $ 2\pi {\bm A(\omega)} = i(\Gb^r(\omega)-\Gb^a(\omega))$.  Finally, the concurrence 
$\mathcal{C}$ is computed 
starting from the expressions for $\langle \hat{N}_\ell \rangle$ ($\ell=L,R$), directly obtained from
Eq.~\eqref{eq:density}, and from $\langle \hat{N}_L \hat{N}_R \rangle$. The latter quantity 
is obtained according to a prescription introduced in Ref.~\cite{Puig_von_Friesen_2011}.

We start with the defining  expression for $\Sigma$. This
in standard NEGF notation on the Keldysh contour reads
\begin{eqnarray}
\int d3\ \Sigma(1, 3) G(3, 2) = -i \int d3\ \nu(1, 3) G_2(1, 3^+; 2, 3^{++}),\nonumber\\\label{eq:sigma}
\end{eqnarray}
in terms of the two-particle Green's function $G_2$:
\begin{equation}
 G_2(1, 2; 3, 4) = - \langle \Psi | \mathcal{T} \left[\hat{c}(1)\hat{c}(2)\hat{c}^\dagger(4)\hat{c}^\dagger(3)\right] | \Psi \rangle. \label{G2}
\end{equation}
Here and in Eq. (\ref{eq:sigma}), $1\equiv (\ell_1,z_1)$ (and similarly for 2, 3, 4), with $\ell_1$ a site index and $z_1$ a time variable
on the Keldysh contour.

We now make the site indices explicit in Eq.~\eqref{eq:sigma}, i.e., $1\rightarrow(\ell, z),\ 2\rightarrow(\ell', z'),\ 3\rightarrow(\bar{\ell}, \bar{z})$,
and set $(\ell', z') = (\ell, z^+) = (L, z^+)$. Furthermore, with nonlocal interactions, the interaction term becomes
$\nu_{\ell\bar{\ell}} (z,\bar{z})= U (\delta_{\ell L} \delta_{\bar{\ell}R} +
\delta_{\ell R} \delta_{\bar{\ell}L}) \delta(z-\bar{z})$. Equation~\ref{eq:sigma} then becomes
\begin{eqnarray}
 \nonumber &\sum_{\bar{\ell}} \int d\bar{z}\ \Sigma_{L\bar{\ell}}(z, \bar{z})G_{\bar{\ell}L}(\bar{z}, z^+)  \\ 
    &= -i U G_{LRLR}(z, z^+; z^+, z^{++}),
\end{eqnarray}
where, here and in the following, $\bar{\ell}=L,R$ in the sums.
Then, using the definition of $G_2$ in Eq.~\eqref{G2},
\begin{eqnarray}
&&\sum_{\bar{\ell}} \int d\bar{z}\ \Sigma_{L\bar{\ell}}(z, \bar{z})G_{\bar{\ell}L}(\bar{z}, z^+) \nonumber \\ 
    && = i U \langle \Psi | \mathcal{T} \left[\hat{c}_{L}(z)\hat{c}_R(z^+)\hat{c}_R^\dagger(z^{++}) \hat{c}^\dagger_L(z^+)\right] | \Psi \rangle,
 \label{eq:sigmaz}
\end{eqnarray}
and by applying the Langreth rules, we arrive at an expression for physical times:
\begin{eqnarray}
\!\!\!&& \frac{-i}{U} \sum_{\bar{\ell}} \int d\bar{t} \big[\Sigma^<_{L\bar{\ell}}(t, \bar{t})G^a_{\bar{\ell}L}(\bar{t}, t^+) 
+ \Sigma^r_{L\bar{\ell}}(t, \bar{t})G^<_{\bar{\ell}L}(\bar{t}, t^+)\big] \nonumber\\
\!\!\!\!&&= \langle \Psi |\left[\hat{c}_R^\dagger(t^{++}) \hat{c}_R(t^+)\hat{c}^\dagger_L(t^+)\hat{c}_{L}(t)\right]|\Psi \rangle =\langle \hat{N}_R\hat{N}_{L} \rangle.
\end{eqnarray}
Finally, moving to frequency space, we have
\begin{eqnarray}
\langle \hat{N}_{L} \hat{N}_R \rangle = &&\frac{1}{U} \sum_{\bar{\ell}} \int \frac{d\omega}{2\pi i}\ (\Sigma^<_{L\bar{\ell}}(\omega)G^a_{\bar{\ell}L}(\omega) \nonumber \\
&&+ \Sigma^r_{L\bar{\ell}}(\omega)G^<_{\bar{\ell}L}(\omega)).
  \label{eq:nn}
\end{eqnarray}
For computational convenience we separate the self-energy into a correlation part and one coming from the Hartree-Fock interaction:
\begin{eqnarray}
\langle \hat{N}_R\hat{N}_{L} \rangle = (\langle\hat{N}_R\rangle\langle\hat{N}_L\rangle
-\langle\hat{c}_{R}\hat{c}_L\rangle\langle\hat{c}_{L}\hat{c}_R\rangle)  \hspace{2cm} \\ + \frac{1} {U}\sum_{\bar{\ell}} \int
\frac{d\omega}{2\pi i}\ (\Sigma^<_{c,L\bar{\ell}}(\omega)G^a_{\bar{\ell}L}(\omega) \nonumber
 + \Sigma^r_{c,L\bar{\ell}}(\omega)G^<_{\bar{\ell}L}(\omega)),
\end{eqnarray}
where the subscript \textit{c} on the self-energy indicates that it contains only the contribution from correlations.

The applicability of this expression has been examined in Ref. \cite{Puig_von_Friesen_2011}. There it was
shown that some approximate but conserving many-body self-energies (among them, those obtained 
in the second Born approximation) do not always guarantee a non-negative value for local double occupancies 
$\langle \hat{N}_{\ell \uparrow} \hat{N}_{\ell\downarrow} \rangle$ in a spinful model with Hubbard interactions.
Here, by contrast, we are considering a spinless model, with nonlocal interactions and correlations
 and in all applications of Eq.~\eqref{eq:nn} in the present paper, no negative correlations were encountered.

\section{Operational nonclassicality}
\label{app:opnonclass}
Following Ref.~\cite{Brask_2015}, we consider the usefulness of the DQD steady state for Bell nonlocality and quantum teleportation.
\subsection{Nonlocality}
\label{App:Nonlocality}

In the CHSH scenario~\cite{Clauser} two players that share a bipartite quantum state, make measurements $x$ and $y$, and obtain outcomes $a$ and $b$. The conditional probability $p(a, b| x, y)$ is nonlocal if it does not admit a local hidden-variable model
\begin{equation}
    p(a, b| x, y) = \sum_{\lambda}p(\lambda) p(a|x, \lambda) p(b|y, \lambda) 
\end{equation}
for any $\lambda$ with an underlying distribution $p(\lambda)$ \cite{Bell}.
In the CHSH scenario $a, b, x, y \in \{0, 1 \}$, and $p(a, b| x, y)$ can be explained by a local hidden-variable model whenever \cite{Clauser}
\begin{equation}
    \text{CHSH} = \sum_{a, b, x, y} (-1)^{a + b + xy} p(a, b| x, y) \leq 2.
\end{equation}

For the density matrix of the double quantum dot, given in Eq.~\eqref{eq:Density Matrix}, the maximal value for the CHSH quantity is given by \cite{Brask_2022}
\begin{equation}
     \text{CHSH} 
    = 2 \sqrt{8 |\alpha|^2 + (1 -  \Delta )^2 - \min \{4 |\alpha|^2,  (1 -  \Delta )^2\} },
\end{equation}
where $\Delta = 2p_0 + 2 p_D$.

For noninteracting electrons, we examine the maximum possible CHSH in the regime $eV \to \infty$, where the corresponding Fermi distributions reduce to $n_L \to 1$ and $n_R \to 0$. Using the expressions for the coefficients of the density matrix, given in Eq.~\eqref{eq:Coefficients local}, we do not find any parameters of the system that give rise to a violation of CHSH inequality.

In the presence of Coulomb interactions, we proceed similarly, and consider the regime described in Sec.~\ref{Sec:Nonlocality}. Under this set of conditions, the probability of the double occupation $p_D$ vanishes, and the maximal value of the CHSH quantity reduces to
\begin{equation}
    \text{CHSH} = 2 \sqrt{8 |\alpha|^2 + (1 - 2p_0)^2 - \min \{4 |\alpha|^2,  (1 - 2p_0)^2\} },
\end{equation}
where the corresponding $|\alpha|$ and $p_0$ are given by Eqs.~\eqref{Eq:CohUinf} and~\eqref{eq: coefficients Semi-local}, respectively. We find that the maximal value of the CHSH quantity is $\text{CHSH} = 2 \sqrt{3/2}$. It is achieved when $\gamma_R/g = 2 \sqrt{2}$ and $\gamma_L/g \to \infty$. 

\subsection{Quantum teleportation}
\label{App:Teleportation}

Quantum teleportation \cite{Bennet_1993} is a protocol where an unknown quantum state is sent between two parties by the means of a shared entangled state and classical communication. In the original implementation, a maximally entangled Bell state $|\psi^-\rangle = (|01\rangle - |10 \rangle)/\sqrt{2}$ was used, giving the average fidelity of teleportation $f = 1$. However, when the parties implementing the protocol do not share a Bell state, but any other less entangled state $\hat{\rho}$, the fidelity of the teleportation is lower with the threshold value $f = 2/3$ for separable states. The average fidelity
$f = (1 + 2F) /3$ can be computed from the so-called singlet fraction \cite{Horodecki_1999}
\begin{equation}
    F = \max_{U} \langle \psi^- | (U \otimes I)~\hat{\rho}~(U \otimes I) |\psi^-\rangle,
\end{equation}
where the optimization is performed over the unitary transformations $U$.
For the density matrix of the double quantum dot, given in Eq.~\eqref{eq:Density Matrix}, 
we have that if its singlet fraction exceeds that of any separable state, i.e., $F >1/2$, then it is given by \cite{Brask_2022}
\begin{equation}
    F = |\alpha| + \frac{1 - p_0 - p_D}{2}.
\end{equation}

In the absence of Coulomb interactions $F$ is saturated at the maximal bias, i.e., when $eV \to \infty$. Using the expressions for the steady state of the density matrix, given by Eq.~\eqref{eq:Coefficients local}, we find the corresponding singlet fraction,
\begin{equation}
    F = \frac{\gamma_L \gamma_R \left(8g^2 + 4g (\gamma_L + \gamma_R) + (\gamma_L + \gamma_R)^2 \right)}{2 (\gamma_L + \gamma_R)^2 (4g^2 + \gamma_L \gamma_R)}.
\end{equation}
It is maximized for the symmetric coupling strength to the baths that fulfills $g/\gamma = (\sqrt{5}-1)/4$, resulting in $F = (3 + \sqrt{5})/8 \approx 0.65$.

Under the conditions described in Sec.~\ref{Sec:Nonlocality} for interacting electrons, $p_D = 0$, whereas $|\alpha|$ and $p_0$ are given by Eqs.~\eqref{Eq:CohUinf} and~\eqref{eq: coefficients Semi-local}, respectively. The singlet fraction is then given by
\begin{equation}
    F = \frac{(8g^2 + 4g \gamma_R + \gamma_R^2)\gamma_L}{2 \gamma_R^2 \gamma_L + 8g^2(\gamma_R + 2 \gamma_L)}.
\end{equation}
$F$ is maximized when $\gamma_R/g = 2 \sqrt{2}$ and $\gamma_L \to \infty$, which coincides with the parameters that saturate concurrence, and amounts to $F = (2 + \sqrt{2})/4 \approx 0.85$. If the coupling strengths to the baths are equal, the maximal singlet fraction occurs at $\gamma/g = 1 + \sqrt{13}$ and is equal to $F = (5 + \sqrt{13})/12 \approx 0.72$.

\bibliography{refs}

\end{document}